\documentstyle[epsfig,12pt]{article}
\newcommand {\bean}  {\begin{eqnarray*}}
\newcommand {\eean}  {\end{eqnarray*}}

\def\Rset {{\rm I \kern-.2em R}} 

\newcommand {\bce}  {\begin{center}}
\newcommand {\ece}  {\end{center}}
\newcommand {\be}   {\begin{equation}}
\newcommand {\ba}   {\begin{array}}
\newcommand {\bea}  {\begin{eqnarray}}
\newcommand {\bfi}  {\begin{figure}}
\newcommand {\ee}   {\end{equation}} 
\newcommand {\ea}   {\end{array}}
\newcommand {\eea}  {\end{eqnarray}}
\newcommand {\efi}  {\end{figure}}

\newcommand {\noi}  {\noindent}

\newcommand {\UNIV}   {Universit\`a }

\def\Rset {{\rm I \kern-.2em R}} 
\def\mathbbH {{\rm I \kern-.2em H}} 
\def\mathbbC {{\rm I \kern-.6em C}} 

\begin{document}
%
\title{
Experimental assessment of a new form of scaling law for near-wall turbulence
}
\author{
B\@. Jacob
\thanks{Dip. Mecc. Aeron., \UNIV di Roma ``La Sapienza'',
        via Eudossiana 18, 00184, Roma, Italy.},
A\@. Olivieri
\thanks{INSEAN, Italian ship model basin, 
        via di Vallerano 139, 00128, Roma, Italy.}
\&
C\@.M\@. Casciola$^*$.}
\maketitle

\section{Abstract}

Scaling laws and intermittency in the wall region of a turbulent flow 
are addressed by analyzing moderate Reynolds number data obtained by single component hot 
wire anemometry in the boundary layer of a flat plate.
The paper aims in particular at the experimental validation of a new form of refined 
similarity recently proposed for the shear dominated range of turbulence, where the classical
Kolmogorov-Oboukhov inertial range theory is inappropriate.
An approach inspired to the extended self-similarity allows for the extraction of the 
different power laws for the longitudinal structure functions at several wall normal distances.
A double scaling regime is found in the logarithmic region, confirming previous 
experimental results.
Approaching the wall, the scaling range corresponding to the classical cascade-dominated range
tends to disappear and, in the buffer layer, a single power law is found to describe
the available range of scales. 
The double scaling is shown to be associated with two different forms of refined similarity.
The classical form holds below the shear scale $L_s$. 
The other, originally introduced on the basis of DNS data for a turbulent channel, is 
experimentally confirmed to set up above $L_s$.
Given the experimental difficulties in the evaluation of the instantaneous dissipation rate,
some care is devoted to check that its one-dimensional surrogate does not bias the results.
The increased intermittency as the wall is approached is experimentally found entirely 
consistent with the failure of the refined Kolmogorov-Oboukhov similarity and the
establishment of its new form near the wall.
\section{Introduction}

Self-similarity is a well explored topic in homogeneous isotropic turbulence.
From Kolmogorov '41 theory onwards \cite{kolm_41}, these flows are expected 
to be scale-invariant in the inertial range, i.e. where the direct effect of 
viscous dissipation is negligible and the dynamics is basically inviscid.
In this range, the scale invariance of the structure functions is rendered as 
a power law with exponent $\zeta(p)$, a nonlinear convex function of the order 
of the structure function itself \cite{frisch}.
The anomalous correction $\tau(p/3)$ to the dimensional prediction accounts 
for intermittency and is strictly related to the statistical properties of the
dissipation field by the refined Kolmogorov similarity hypothesis 
\cite{kolm_62}, \cite{Oboukhov}. 
The ideas behind this theory have been the subject of close scrutiny, see e.g. 
Chen {\sl et al.} \cite{Kraichnan} and Wang {\sl  et al.} \cite{Brasseur} where a 
detailed analysis of the dissipation field is performed,
and the refined similarity may now be considered a well assessed feature of homogeneous 
isotropic turbulence.
The exponents are known with good accuracy, and the procedure for their evaluation has been 
improved by the introduction of the extended self-similarity (ESS) by Benzi {\sl et al.} 
\cite{benzi_1}. 
The success of homogeneous isotropic theory is motivated by its ability in describing certain 
universal properties of turbulence which are presumably recovered also in complex flows at 
sufficiently fine scales, see e.g. the experimental data of Saddoughi and Veeravalli 
\cite{saddoughi} in the logarithmic region of a turbulent boundary layer.

In the present paper, we address the issue of self-similarity and intermittency in a 
turbulent boundary layer, by discussing experimental results obtained by hot wire 
anemometry.
In addition to the two lengths which are typical of homogeneous and isotropic turbulence, 
namely the dissipation and the integral scale, the mean shear introduces as a further 
characteristic quantity the shear scale \cite{Hinze}. 
As we shall see, depending on the distance from the wall, the shear scale may change 
appreciably, from values close to the viscous scale in the  buffer layer up to the integral 
scale in the higher part of the logarithmic region \cite{Amati}.

Physically, as described with a few more details in section~III, 
the mean shear alters the mechanism of energy transfer through the scales of 
turbulence by  originating a continuous injection of energy. 
The production mechanism is active at scales larger than
the shear scale, where the classical form of scaling law described by the 
refined Kolmogorov-Oboukhov theory may be expected to fail.
This effect was clearly found in the data from a DNS of a channel flow where the ESS scaling 
exponents of the longitudinal structure functions depended on the wall normal 
distance \cite{PRL}. These observations were successfully explained in terms of a new form
of refined similarity in \cite{PF} (see also \cite{cercignani} for a complete 
theoretical presentation).
All these findings have been confirmed by a successive DNS of homogeneous shear
flow \cite{paolino}, undertaken almost in parallel with the present 
experimental investigation.

From the experimental point of view, scaling laws in shear flows have been already considered, 
see  \cite{Antonia1}. Concerning wall turbulence, estimates for the effective ESS exponents
 as a function of wall distance may be found in \cite{Antonia2}, \cite{Onorato}. 
Recently, a double scaling regime was clearly detected by Ciliberto and coworkers \cite{ciliberto} 
in the logarithmic region of a turbulent boundary layer.
Here we confirm the above results and add more information on the behavior
of turbulence in the lower part of the log-region and in the
buffer, were a unique power law is found throughout the entire range
of scales.
As a major contribution, we show that the two scaling regimes can be understood
in terms of two different forms of scaling law, namely the refined Kolmogorov-Oboukhov 
similarity \cite{Monin}, for separations below the shear scale, and the new form of 
refined similarity \cite{PF} above the shear scale.

Technically, the main difficulty in assessing similarity laws which involve 
statistics of the dissipation field is the resolution required to follow the entire
range of scales down to the Kolmogorov length.
This is not easily accomplished in a laboratory experiment, essentially due to 
the transverse dimension of the probe which acts as a filter on the smallest 
scales.
Moreover, to properly evaluate the dissipation, one should have access to the whole
velocity gradient tensor. Instead, we can measure only the streamwise 
velocity at a fixed position as a function of time. As costumery, by invoking 
Taylor hypothesis of frozen turbulence, we are able to obtain a single 
component of the velocity gradient, namely the derivative of the
streamwise velocity in the mean flow direction. 
Hence the need to use a surrogate of the real dissipation which, however, is 
in principle quite rough an approximation.
To corroborate the results obtained by the surrogate, we account for a conjectured behavior 
of the dissipation field based on DNS data (\cite{PF}, \cite{paolino}), suggesting the 
robustness of the observed scalings.

The paper is organized as follows. In section~II a brief description of the experimental 
set-up is provided and the basic dimensionless parameters and the global 
statistical features of shear turbulence are briefly discussed.
Section~III concerns scaling laws for the structure functions and presents the two forms of
refined similarity.
Section~IV is devoted to the experimental evaluation of the two refined similarity theories
and discusses their relationship with the intermittency of the wall region.
Final comments and concluding remarks are reported in section~V.
\section{Global features of the flow}

The data we are going to discuss are obtained from a zero pressure gradient boundary
layer on a flat plate $1200 \, {\rm mm}$ long positioned in the test section of a wind tunnel 
operated at a free stream velocity $U_{\infty}$ of  $12 \, {\rm m/s}$. 
Measurements were taken $750 \, {\rm mm}$ downstream the leading edge of the plate, where the 
local thickness of the boundary layer $\delta$ was about $\simeq$ $25 \, {\rm mm}$.
The Reynolds number, based on momentum thickness, was $Re_{\theta} \simeq 2200$.

A constant temperature anemometer was used to acquire  the streamwise velocity and the 
signal was low-pass filtered at $8000 \, {\rm Hz}$ and successively sampled at 
$75000 \,  {\rm samples/s}$. 
The typical length of each record was $50 \, {\rm s}$.  The friction velocity $u_{\tau}$, 
evaluated via Clauser plot, is estimated as $u_{\tau} = .489 \, {\rm m/s}$. 

The plot on top of figure~\ref{mean_profile} gives the mean velocity profile 
$u^+ = u/u_{\tau}$ as a function of the distance from the wall in inner units,
$y^+ = y u_{\tau} /\nu$, where $\nu$ is the kinematic viscosity. 
The classical logarithmic law, given by the straight line in the log-log plot,

\begin{eqnarray}
\label{log_law}
u^+  & = & \frac{1}{k} \, \ln y^+ \, + \, C \, ,
\end{eqnarray}

\noi corresponds to $k=0.40$ and $C=5.5$. Near the wall the data approach the linear behavior 
typical of the viscous sublayer.
The plot on the bottom of the figure gives the root mean square value of the fluctuating 
velocity. 

Concerning the dynamics of turbulent fluctuations, the wall region of the boundary layer is 
characterized by two basic mechanisms, the process of energy cascade towards small scales 
and the production of turbulent kinetic energy associated to the presence of the shear.
Their relative importance may be estimated in terms of two dimensionless parameters involving
the shear length scale $L_s$ \cite{Hinze},

\begin{eqnarray}
\label{Ls}
L_s=\sqrt{\frac{\bar{\epsilon}(y)}{S^3}} \, ,
\end{eqnarray}

\noi where $\bar{\epsilon}(y)$ is the local average of the energy dissipation rate and $S$ the local 
mean shear, $S = \partial u/\partial y$. 
The two dimensionless parameters, expressed as

\begin{eqnarray}
\label{Sc}
S^*_c = S (\nu /\bar{\epsilon})^{1/2}= \left(\frac{\eta}{L_s}\right)^{2/3}  \, , 
\end{eqnarray}

\noi and

\begin{eqnarray}
\label{Sstar}
S^*=\frac{S u_{rms}^2}{\bar{\epsilon}}=\left(\frac{l_d}{L_s}\right)^{2/3}
\end{eqnarray}

\noi  compare the shear scale to the local Kolmogorov length, 
$\eta = \left(\nu^3 / \bar{\epsilon}(y)\right)^{1/4}$, and to a large scale inertial length, 
$l_d=u_{rms}^3 / \bar{\epsilon}$, respectively.

In the log-region, since the production of turbulent kinetic energy, 
$\Pi = - S <u' v'> \simeq S u_\tau^2$, is almost in local balance with the dissipation, we have 
$\bar{\epsilon} \simeq u_\tau^4 /(\nu y^+ k)$, hence 

\begin{eqnarray}
\label{S_Sc_log}
S^* = {\cal O}(1) &  \qquad & S_c^* \simeq 1./(u_\tau y k)^{1/2} \nonumber   \\
 & S^*/S_c^* \simeq (u_\tau y k)^{1/2} \, . & 
\end{eqnarray}
\noi These estimates suggest that increasing the distance from the wall, one should
reach a position where the effect of the shear on turbulent fluctuations becomes
undetectable.  On the opposite, approaching the wall, $L_s$ tends to collapse on $\eta$, and 
the shear affects the fluctuations more and more pronouncedly.


The statistical results we are going to discuss are relative to three measurement 
points: one located within  the buffer layer, and the others in the log-region.
The characteristic parameters at these locations are summarized in Table \ref{table}.

Figure~\ref{spectra} (top) presents two typical one-dimensional 
spectra, in the log-region and in the buffer layer. Indicated are also the two 
slopes, $-5/3$ for the purely inertial range and $-1$ for the energy production range,
respectively, which should be locally recovered.
In fact, no distinct power-laws can be seen in our energy spectra, presumably due to the 
relatively low values of the local turbulent Reynolds number,

\begin{eqnarray}
\label{Re_lambda}
Re_{\lambda} & = & u_{rms} \lambda / \nu \, \,
\end{eqnarray}

\noi where $\lambda = u_{rms}/<(\partial u / \partial x)^2>^{1/2}$ is the local Taylor 
microscale, see table~\ref{table}.
Nonetheless, a clear trend towards 
the formation of a $k^{-1}$ range is observed as the wall
is approached.

\noi This is consistent with the results  presented in
\cite{ciliberto}, where a clear scaling emerged and the characteristic slopes were found to 
evolve from $-5/3$ towards $-1$ when reducing the distance from the wall. 
To conclude this preliminary analysis, figure~\ref{spectra} (bottom) reports the dissipation 
spectra.
\section{Similarity laws for shear dominated turbulence}

The statistical features of turbulence are best addressed by considering the
longitudinal increments of the streamwise velocity component, 
\begin{eqnarray}
\delta V \, = \,  u(x + \ell ,y,z,t) - u(x,y,z,t)  \, .
\label{V_increment}
\end{eqnarray}

\noindent For a boundary layer, the longitudinal velocity increment is a random function which, 
neglecting the weak effect of non-parallelism, may be considered homogeneous both in the 
streamwise and the spanwise direction, $x$ and $z$ respectively, and in time. 
Hence the pdf of $\delta V$ depends 
\noi only on the separation, $\ell$, and the wall normal distance, $y$.
Typical examples are given in figure~\ref{pdf} which shows the pdf for two 
separations at two different wall distances.
As the separation is decreased, the tails of the pdf rise and this effect is more pronounced 
in the buffer region.    
This behavior is associated with the intermittency of the velocity increments, as expressed 
in terms of the flatness \cite{frisch},
\begin{eqnarray}
F_4(\ell, y) \, = \, <\delta V^4 > / <\delta V^2 >^2 \, ,
\label{flatness}
\end{eqnarray}
and corresponds to an increased intermittency as the wall is approached.

\subsection{Structure functions}

The pdf of the velocity increment is characterized by the structure functions, i.e. its moments 

\begin{eqnarray}
S_p(\ell,y) \, = \, <|\delta V|^p> \, ,
\label{Structure}
\end{eqnarray}

\noindent or, in wall units, $S_p^+(\ell^+,y^+) \, = \, S_p/u_{\tau}^p \, $.
In our case, the structure functions depend on the separation $\ell$, the wall normal coordinate $y$
and, in general, on the Reynolds number $Re_\theta = u_\tau \, \theta/ \nu$, where $\theta $ is 
the momentum thickness.

For fully developed homogeneous isotropic turbulence, the structure functions 
are given as power laws of separation
\cite{frisch}
\begin{eqnarray}
S_p(\ell) \, \propto \ell^{\zeta_p} \, ,
\label{Scaling_HI}
\end{eqnarray}

\noi where the scaling exponents differ from the dimensional prediction, $p/3$, by the 
anomalous correction $\tau(p/3)$, i.e. $\zeta_p   =  p/3 \, + \, \tau(p/3)$.
Since the exponent is a convex function of the order $p$, the flatness 
increases as the separation is decreased in the inertial range \cite{frisch},
\begin{eqnarray}
 F_4(\ell) & \propto  & \ell^{\, \tau(4/3) \, - \, 2 \, \tau(2/3)} 
\, ,
 \label{intermittency_HI}
\end{eqnarray}

\noi consistently with the intermittency of the velocity increments.

Scaling laws in terms of separation can hardly be detected at laboratory scale, 
for the insufficiently large Reynolds number available.
This difficulty has been recently overcome by Benzi and coworkers \cite{benzi_1}, who
proposed a relative form of scaling, the extended self-similarity (ESS), where the third order 
structure function is assumed as independent variable instead of the separation
\begin{eqnarray}
S_p \, \propto \,  S_3^{\hat{\zeta_p}}.
 \label{Benzi}
\end{eqnarray}
\noi For large Reynolds numbers, since $S_3 \propto \ell$ by the Karman-Howarth equation, 
the relative scaling is a direct consequence of eq.~(\ref{Scaling_HI}). 
As an advantage, the scaling 
range is substantially extended and scaling laws emerge also 
for relatively small Reynolds number.
Clearly, in 
fully developed homogeneous turbulence eqs.~(\ref{Scaling_HI}), (\ref{Benzi}) imply 
$\hat \zeta_p = \zeta_p$.
The relative scaling may also be found under more general conditions than homogeneous 
isotropic turbulence, e.g. in turbulent shear flows, where, however, 
$\hat \zeta_p$  and $\zeta_p$ may be entirely different objects.

Concerning our data, two examples of third order 
structure functions  are plotted in 
figure~\ref{S_3}. No scaling is observed, apart from the linear behavior at small separations.

Figure \ref{ESS} presents similar data in terms of ESS, in particular  
$S_6$ vs. $S_3$.
At $y^+=30$, a single fit, with slope ${{\hat \zeta}_6} \simeq 1.54$,
suffices for the whole range. This value is substantially
lower than $1.78$, expected value for homogeneous and isotropic turbulence.   
The plot for $y^+=70$ manifests instead two distinct regions, each with a 
reasonably well-defined scaling exponent, 1.78 and 1.54 respectively.
This double-scaling regime is also found at $y^+=115$ where the transition between
the two behaviors has moved towards larger separations and most of the range is associated to
the slope $1.78$.

The scale which approximately identifies the crossover between the two 
regimes is proportional to $L_s$, which, in the log region, may be estimated as  
$k\, y \simeq 0.4\, y$.
Below the crossover, the behavior of the structure functions follows the predictions 
for homogeneous and isotropic turbulence ($\hat \zeta_6 = 1.78$), consistently with the
assumption of the distance from the wall as the characteristic separation for its validity
\cite{Hinze}.

\subsection{Refined Kolmogorov Similarity}

Physically, intermittency is associated to the spotty nature of the energy dissipation field,
as implied by the refined Kolmogorov similarity hypothesis \cite{kolm_62}, 
\begin{eqnarray}
 \label{K62}
S_p(\ell) & \propto & <\epsilon_\ell^{p/3} > \, \ell^{p/3}
\end{eqnarray}

\noi where $\epsilon_\ell$ is the rate of energy dissipation, $\epsilon_{loc}$, spatially
averaged on a domain of characteristic dimension $\ell$.

In homogeneous isotropic turbulence, the moments of the coarse grained dissipation field, 
$\epsilon_\ell$, scale as $<\epsilon_\ell^{p} >  \propto  \, \ell^{\, \tau(p)}$
consistently with equation~(\ref{Scaling_HI}).
A number of models have been proposed to predict $\tau(p)$, beginning with the original 
log-normal model \cite{Oboukhov} up to the log-Poisson model proposed by She and L\'ev\^eque 
\cite{She} able to accurately fit the experimental data, 

\begin{eqnarray}
\tau_{sl}(p) \, = \, - \frac{2}{3} \, p \, + \, 2 \, \left( \, 1 - (2/3)^p \, \right) \, .
 \label{She_leveque}
\end{eqnarray}

\noi In its extended form, the refined similarity reads \cite{benzi_2}

\begin{eqnarray}
 \label{RKSH_old}
S_p(\ell) & \propto & <\epsilon_\ell^{p/3} > \, S_3(\ell)^{p/3} \, .
\end{eqnarray}

Following the discussion of section~III.A, for the wall region of the boundary layer we expect 
no substantial alteration of the scaling laws for separations smaller than a scale of the 
order of the shear length $L_s$.
When the separation becomes larger than the shear scale, the mechanism of energy transfer is affected by 
the production of turbulent kinetic energy.
This reasoning recently suggested a different form of refined similarity \cite{PF},
\begin{eqnarray}
 \label{RKSH_new}
S_p(\ell) & \propto & <\epsilon_\ell^{p/2} > \, S_2(\ell)^{p/2} \, ,
\end{eqnarray}

\noi expected to replace eq.~(\ref{RKSH_old}) in the range of scales 
$L_s << \ell << L$, where $L$ is the integral scale.

For a direct assessment of eqs.(\ref{RKSH_old}), (\ref{RKSH_new}) the dissipation field should 
be available, an issue to be considered in more detail in the next section. 
Here we note that both equations may be rewritten as

\begin{eqnarray}
 \label{GESS}
S_p/S_{\alpha}^{p/{\alpha}} & \propto & <\epsilon_\ell^{p/{\alpha}}>\qquad\qquad\alpha =3,2\,.
\end{eqnarray}

\noi In eq.~(\ref{GESS}) the structure functions are grouped together on the left hand side, 
set apart from the moments of the dissipation written on the right. 
The left hand side can be evaluated directly from the velocity signal with no additional 
assumption besides Taylor's hypothesis. 
These quantities are plotted as functions of $S_3$ in figure \ref{verifica_refined}, for both 
$\alpha=3$ and $\alpha=2$, top and bottom, respectively. 
Concerning the measurement point in the buffer (circles), a single power-law is observed in 
both cases.
When moving to points in the log-layer, this representation confirms the coexistence of two 
distinct power-laws, as already found from the analysis of the structure functions. 
The crossover between the two ranges occurs at $\ell \propto L_s$, where the proportionality 
factor roughly corresponds to $10$, as already found in \cite{ciliberto}.
Typically, since the shear length increases with the distance from the wall, the 
relative extension of the two scaling ranges is different. 
Nonetheless the exponents remain rather constant, thus
suggesting a certain universality of the two scaling regimes, as seen by the comparison of the
data at $y^+=70$ and $115$, diamonds and triangles, respectively.
Note that the abscissa $S_3$ in the plot has been rescaled to achieve collapse of the curves
at large separations.

Concerning the shear-dominated range at larger separations, in the log-layer we recover
the same slope found in the buffer. 

From the top plot of figure~\ref{verifica_refined}, in the log-layer and for the 
cascade-dominated range at smaller separations, we find the slope $s_3\simeq-.22$, see the 
dotted lines in the figure.
Concerning the scaling range at larger separations, from the bottom plot of 
figure~\ref{verifica_refined} we estimate the slope $s_2 \simeq -.59$, solid line in the figure.
For wall turbulence, the interpretation of either $s_3$ and $s_2$ in terms of possible scaling 
laws for the dissipation field is not obvious and will be discussed further in the next section.
We only anticipate here that the value of $s_3$ is essentially the same well known from 
homogeneous and isotropic turbulence, where the Kolmogorov-Oboukhov theory holds, and $s_3$ 
represents the scaling exponent of $<\epsilon_\ell^{2}>$ with respect to $S_3$.

To summarize, the present results suggest that turbulence in the near-wall region is 
characterized by the juxtaposition of {\em{only}} two basic scaling regimes. 
As will be shown in the next section, our data support the conclusion that these two regimes 
correspond to the classical and the revised form of similarity, eqs.~(\ref{RKSH_old}) and 
(\ref{RKSH_new}), respectively.

\section{Experimental evaluation of the refined similarity laws}
In order to provide direct evidence of the two forms of refined similarity conjectured to be
the origin of the scaling laws described in the previous section, experimental data for the 
instantaneous dissipation are needed.
Actually, single component hot-wire anemometry gives access only to the streamwise velocity component
at a fixed position as a function of time. In these conditions, we have to use a surrogate for 
the dissipation field to estimate $\epsilon_{loc}$ from $\partial u/\partial t$.

To this purpose we replace the actual dissipation, $\epsilon_{loc}$, with its
one-dimensional isotropic version, $15 \, \nu  \, (\partial u / \partial x)^2 $, where the
x-derivative is estimated from the time signal via the Taylor hypothesis of frozen turbulence.
To account for the presence of the mean shear, we introduce the additional term $\nu S^2$,
hence we roughly replace the instantaneous dissipation with

\begin{eqnarray}
\epsilon' & = & \nu \left( \, S^2 \, + \, 
           15 \, \frac{\partial u}{\partial x} \, \frac{\partial u}{\partial x} \, \right) \, .
\label{surrogate}
\end{eqnarray}

\noi Introducing this surrogate, the classical and new form of scaling laws become 

\begin{eqnarray}
 \label{Sur_RKSH_new_old}
 \frac{S_p(\ell) } { S_\alpha(\ell)^{p/\alpha} }
 & \propto &  
<{\epsilon'}_\ell^{p/\alpha}> \,  \qquad \qquad \alpha = 3,2  \, ,
\end{eqnarray}

\noi where ${\epsilon'}_\ell=\frac{1}{\ell}\,\int_{-\ell/2}^{\ell/2}\,\epsilon'\,dx$, 
and $\alpha=3, 2$ gives the classical and the revised version of the refined similarity,
respectively.

\subsection{Refined similarity and dissipation}

The left hand side of eq.~(\ref{Sur_RKSH_new_old}) is plotted vs $S_3$ in 
figure~\ref{dissipazione_sorrogato_refined}, where the filled symbols correspond to $\alpha=3$
(classical form) and the open symbols to $\alpha=2$ (new form).
The three measurement points are separately analyzed, from top to bottom 
$y^+=30, \, 70, \, 115$. 
The symbols are a rearrangement of the curves appearing in figure~\ref{verifica_refined}. 
Here the figure is centered on the
dotted and the solid curves, which, for each data point, represent
the right hand side of eq.~(\ref{Sur_RKSH_new_old}), namely the moments of the dissipation
as estimated by the surrogate (\ref{surrogate}).
In each of the three cases, the dotted line corresponds to $\alpha=2$, the solid
line to $\alpha = 3$.

In the buffer, top plot, the classical form of refined similarity ($\alpha=3$) is manifestly
violated, as implied by the mismatch between the filled circles ($S_6/S_3^2$) and the
solid line ($<{\epsilon'}_\ell^2>$).
Instead, the new form ($\alpha=2$) adapts to the data with remarkable accuracy, as suggested
by the comparison of the open circles ($S_6/S_2^3$) with the dotted line 
($<{\epsilon'}_\ell^3>$).
In physical terms, we find that at $y^+=30$ the cascade-dominated dynamics described by the 
Kolmogorov-Oboukhov theory is not recovered and, far from the purely dissipative region, the 
entire dynamics is well captured by the new refined similarity.

This result should be compared with the picture emerging from the plot in the middle, 
corresponding to $y^+=70$. 
There we find two ranges. One at larger separations, where the dotted line
matches the open diamonds. The other where the solid curve falls on top of
the filled diamonds. 
The first behavior is exactly the same as discussed for the buffer implying the 
establishment of the new refined similarity at large scales.
The other corresponds to the Kolmogorov-Oboukhov refined similarity which is recovered  at 
scales sufficiently smaller than $L_s$ and well above $\eta$. 
This double scaling regime is found also at $y^+=115$, bottom plot where 
the extension of the classical cascade-dominated range is increased and that of the 
shear-dominated range is reduced.

Let us restate the main results of this section to underline their physical interpretation.
On the one hand, in the log layer, we find the coexistence of the classical and new form of 
refined similarity. 
On the other hand, as the wall is approached, the extension of the classical range is reduced
to finally disappear when approaching the buffer layer where only the shear-dominated
range corresponding to the new refined similarity remains.

Figure~\ref{surrogate_refined} confirms this conclusion, by presenting the same data
in a compensated form, i.e by plotting along the ordinates the left hand sides of 
equations~(\ref{Sur_RKSH_new_old}) divided by the respective right hand sides.
A clear plateau is described by the filled symbols at smaller separations in the log-region
(classical refined similarity).
A less extended scaling region is found at larger separations in the curve indicated by the 
open symbols (the new refined similarity in the shear-dominated range).
In the buffer, only the open circles describe an almost straight horizontal line,
to show that the entire range is filled by the new form of scaling.
To avoid possible misunderstandings, in this figure we have plotted the data 
vs the geometrical separation $\ell$, to evidence the physical interpretation of the results
we have been discussing so far.

In conclusion, despite our use of the surrogate (\ref{surrogate}) for the dissipation, these 
results show that the two behaviors described in the previous section are well interpreted in 
terms of  the two forms of refined similarity.
It would however be advisable to show that we are not biased by the strong 
limitations introduced by the surrogate.
This could be done by replacing the measured moments of the dissipation with suitable 
theoretical results.
If we consider for the time being our results as unaffected by the use of the surrogate, the 
scaling exponent $s_3$ (see section~III.B) found in the cascade-dominated range should be 
interpreted as the scaling exponent of the second moment of the coarse grained dissipation 
field with respect to $S_3$.
We know that its value is in agreement with that known for homogeneous isotropic
turbulence.
Moreover if we assume eq.~(\ref{RKSH_new}) to give the correct description of turbulence in the 
shear-dominated range, $s_2$ should also be interpreted as the scaling exponent of the second 
moment of the dissipation with respect to $S_3$.
The analysis of DNS data for the channel flow \cite{PF} and the homogeneous shear flow 
\cite{paolino} support the conjecture that the scaling properties of the dissipation field 
are, to a first instance, unaltered by the shear.
If this is the case, the scaling exponent of the third order moment of the dissipation with 
respect to $S_3$ should be close to the value $\tau_{sl}(3)=.592$ as evaluated from the
She-L\'ev\^eque model, eq.~(\ref{She_leveque}).

Returning to figure (\ref{verifica_refined}), the solid line superimposed to the plots on the 
bottom has precisely the slope $-.592$. 
The line fits almost exactly the experimental results for the surrogate-independent combination
of structure functions $S_6/S_2^3$ in the shear-dominated range.

This argument is strongly in favor of the conclusion that the 
surrogate, though not accurate, does not obscure our physical interpretation: 
near wall turbulence is characterized by two different scaling regimes, one for the 
cascade-dominated range, the other for the shear-dominated scales. 
The existence of the two scalings is associated with two different forms of refined 
similarity, the Kolmogorov-Oboukhov similarity for the cascade-dominated regime, its revised 
form (\ref{RKSH_new}) for the shear dominated regime \cite{PF}.
The cross-over between the two regimes takes place rather abruptly, implying that two 
radically different dynamics take over in the two distinct ranges. 
We mention, in passing, that the two scaling regimes we have presently identified could be 
described, in principle,  by a unified form of generalized structure function, as recently 
proposed in \cite{Toschi} (see also \cite{cercignani}).

\subsection{Near-wall intermittency and dissipation}

Let us now return to the issue of intermittency.
Recent work on scaling laws for near-wall turbulence has pointed out that the amount of 
intermittency increases as the wall is approached from the log-layer towards the buffer.
This effect may be appreciated in figure~\ref{flat_exp}, where the flatness of the velocity 
increments is reported for two measurement points, one  in the buffer, the other in the log 
layer.
Clearly, when the separation decreases, the flatness factor is seen to increase by far more
in the case of the point closest to the wall, giving a quantitative measure of the larger 
intermittency of the velocity increments.

In the context of the classical form of refined similarity this should imply a substantial alteration of the 
statistical properties of the dissipation field.
Actually, from equation~(\ref{RKSH_old}) the flatness of the velocity increments, 
eq.~(\ref{flatness}), can be expressed as 
\begin{eqnarray}
F_4(\ell, y) & = & \frac{ <\epsilon_\ell^{4/3} > } { <\epsilon_\ell^{2/3} >^2 }\, . 
\label{flatness_old} 
\end{eqnarray}

\noi This expression is expected to be able to describe the intermittency in those regions
where the classical scaling law exists, i.e. in the range below the shear length $L_s$ 
and above the Kolmogorov length $\eta$.
As an example, we have found that at $y^+ = 115$ most of the range of scales available in our 
experiments is characterized by scaling laws corresponding to the classical refined similarity.
Actually, as shown in figure~\ref{flat_exp}, eq.~(\ref{flatness_old}) is able to reproduce 
with reasonable accuracy both the quantitative and qualitative behavior of the flatness factor,
as shown by the filled triangles, eq.~(\ref{flatness_old}), which follow the dashed line giving 
the flatness in terms of its definition (\ref{flatness}).
However when we consider the data pertaining to $y^+=30$, eq.~(\ref{flatness_old}) is clearly 
seen to fail. 
This comes as no surprise, since it has been already shown that close to the wall the 
classical form of similarity itself breaks down.
As we have verified, at $y^+=30$ the new form of scaling described by eq.~(\ref{RKSH_new}) 
is established in almost the entire range of scales above $\eta$. 
This allows us to express the flatness factor as

\begin{eqnarray}
F_4(\ell, y) & = & \frac{ <\epsilon_\ell^{2} > } { \bar \epsilon^2 }\, . 
\label{flatness_new} 
\end{eqnarray}

\noi In the figure, this estimate for the flatness is compared with the values directly 
evaluated from the velocity increments, open circles to be compared with the solid line,
where it is clearly seen to capture with reasonable accuracy the amount of intermittency 
found in the velocity increments near the wall.

As commented in \cite{PF}, the ability of the new form of scaling law to capture the amount 
of intermittency is an important indication of its validity, since it entails the physical 
interpretation of the increased intermittency found by a number of investigators in the near 
wall region.
Actually, as we already mentioned, the statistical properties of the dissipation field
eq.~(\ref{surrogate}) do not seem to change significantly from the log-region to the buffer. 
In fact, intermittency increases because of the changing balance between energy 
transfer and dissipation originated by the large production which occurs locally in 
wavenumber space at shear-dominated scales.
\section{Concluding remarks}

We have discussed the scaling properties of the longitudinal velocity increments in a 
turbulent boundary layer.
No scaling is apparent in our data when the structure functions are expressed as a function of 
the separation. 
The extended self-similarity, instead, makes clear the existence of a double
scaling regime in the log-region of the boundary layer.
The cross-over between the two regimes is found to be controlled by the shear scale.
Below the cross-over, the values of the scaling exponents are in good agreement with those
of homogeneous isotropic turbulence. 
Since the shear scale is proportional to the wall normal distance, we confirm that a 
substantially isotropic dynamics is recovered at scales smaller than the distance from the
wall, as expected from classical descriptions of wall bounded flows \cite{townsend}.
Above the cross-over, the crucial effect of the shear induces a different form
of power law, with lower values of the exponents.
Only this form of scaling survives when the buffer region is approached, where the classical 
cascade-dominated range disappears.

The values of the exponents found in the shear-dominated range are suggesting an increase of 
intermittency, as confirmed by the analysis of the flatness of the velocity increments.
In fact, no direct conclusion can be drawn only from power laws in terms of the third order 
structure function in wall turbulence. 
Actually, in presence of shear, the Karman-Howarth equation does not provide the necessary
link between $S_3$ and separation.

We find that the cascade-dominated regime in the log-layer is consistent with the classical
form of refined similarity. The anomalous correction to the scaling exponents of the structure 
functions can then be ascribed entirely to the dissipation field, along the line of the
theory of Kolmogorov and Oboukhov.
Instead, in the shear-dominated range and in the buffer-layer the new values for the 
scaling exponents should be understood in terms of the new form of refined similarity
introduced in \cite{PF}.

From our hot-wire data, by using a suitable one-dimensional surrogate of the energy 
dissipation, we find that the increased intermittency is essentially related to the
change in the structure of the similarity law and not to a substantial change in the 
statistical properties of the dissipation field.
Actually, we are able to fit our data by using the new form of refined similarity 
together with the scaling exponents of the dissipation, as known from homogeneous isotropic
turbulence.

Since the scaling properties of the dissipation field are substantially the same as
those of homogeneous isotropic turbulence, we can ascribe the increased intermittency 
to the different structure of the new refined similarity.
This is actually confirmed by directly comparing the measured values of the flatness 
of the velocity increments with those inferred from the new refined similarity, which is 
found able to predict the increased intermittency observed as the wall is approached.

The present experimental investigation, while  supporting the findings of previous DNS of 
channel \cite{PF} and homogeneous shear turbulence \cite{paolino}, clearly show the 
coexistence of the two scaling regimes in the log-layer, a conclusion which could only be 
conjectured on the basis of DNS data, due to insufficient scale separation.

\newpage
\begin{figure} [!hb]
   \caption{Mean velocity profile (top) and $u_{rms}$ (bottom) as 
            a function of the distance from the wall.
            \label{mean_profile}}
\end{figure}
\begin{figure} [!hb]
   \caption{Velocity and dissipation spectra at $y^+=115$ and $y^+=30$
            \label{spectra}}
\end{figure}
\begin{figure} [!hb]
   \caption{Pdf of $\delta V$, at $y^+ = 115$ and  $y^+=30$ (triangles and circles 
            respectively),
            for two separations, $\ell^+ = 28$ (top) and $\ell^+=180$ (bottom).
            The variable has been centered at its mean value and normalized by its 
	    standard deviation.	
            \label{pdf}}
\end{figure}
\begin{figure} [!hb]
    \caption{$S_3^+$ vs $\ell^+$ at $y^+=115$ and $30$, top and bottom respectively.
             \label{S_3} }
\end{figure}
\begin{figure} [!hb]
    \caption{$S_6$  vs. $S_3$ (ESS). From top to bottom:  $y^+=30$, $70$ and $115$. 
             The slopes of the solid and the dashed lines are $1.78$  and $1.54$, 
             respectively.
             \label{ESS}}
\end{figure}
\begin{figure} [!hb]
   \caption{Top: log-log plot of $S_6/S_3^2$ vs $S_3$, $p=6$, $\alpha=3$ in eq.~(\ref{GESS}), 
                                                        at $y^+=115$ (filled triangles),
                                                                $70$ (filled diamonds),
                                                            and $30$ (filled circles). 
            Bottom: log-log plot of $S_6/S_2^3$ vs $S_3$, $p=6$, $\alpha=2$, 
                                                           at $y^+=115$ (open triangles),
                                                                   $70$ (open diamonds),
                                                               and $30$ (open circles). 
            The slopes of the dotted and the solid lines are $s=-.222 $ and $s= -.592$,
            respectively. They are discussed at the end of section~IV.A.
            \label{verifica_refined}}
\end{figure}
\begin{figure} [!hb]
    \caption{Log-log plot of $S_6/(S_3^2 \, \epsilon_\ell^2)$ (filled symbols)
             and of $S_6/(S_2^3 \, \epsilon_\ell^3)$ (open symbols) vs $S_3$, 
             see caption of figure~\ref{verifica_refined}.
             From top to bottom, $y^+=30, \, 70, \, 115$.
             The dotted lines correspond to $<\epsilon_\ell^3>$, the solid line
             to $<\epsilon_\ell^2>$.
             The moments of the dissipation field are estimated from the one-dimensional
             surrogate, eq.~(\ref{surrogate}). The curves have been arbitrarily shifted
             along the ordinate.
             \label{dissipazione_sorrogato_refined}}
\end{figure}
\begin{figure} [!hb]
    \caption{Compensated plot vs separation $\ell$ of $S_6/(S_3^2 \, \epsilon_\ell^2)$ 
             (filled symbols)
             and $S_6/(S_2^3 \, \epsilon_\ell^3)$ (open symbols) at $y^+=115$ and 
             $y^+=30$, upper and lower part respectively.
             The moments of the dissipation field are estimated from the one-dimensional
             surrogate, eq.~(\ref{surrogate}).
             \label{surrogate_refined}}
\end{figure}
\begin{figure} [!hb]
    \caption{Flatness of the velocity increments at $y^+=30$ and $115$, solid and dotted line, 
             respectively.
             The circles correspond to the estimate of the flatness according to the new
             form of refined similarity, eq.~(\ref{flatness_new}) at $y^+=30$.
             The triangles give the flatness as estimated by the classical refined similarity, 
             eq.~(\ref{flatness_old}),
             at $y^+=115$.
             \label{flat_exp}}
\end{figure}
\setcounter{figure}{0}
\newpage
\begin{figure}[b!] 
\begin{center}
{\Large FIGURES}
\end{center}
   \vspace*{2.cm}
   \centerline{ \epsfig{figure=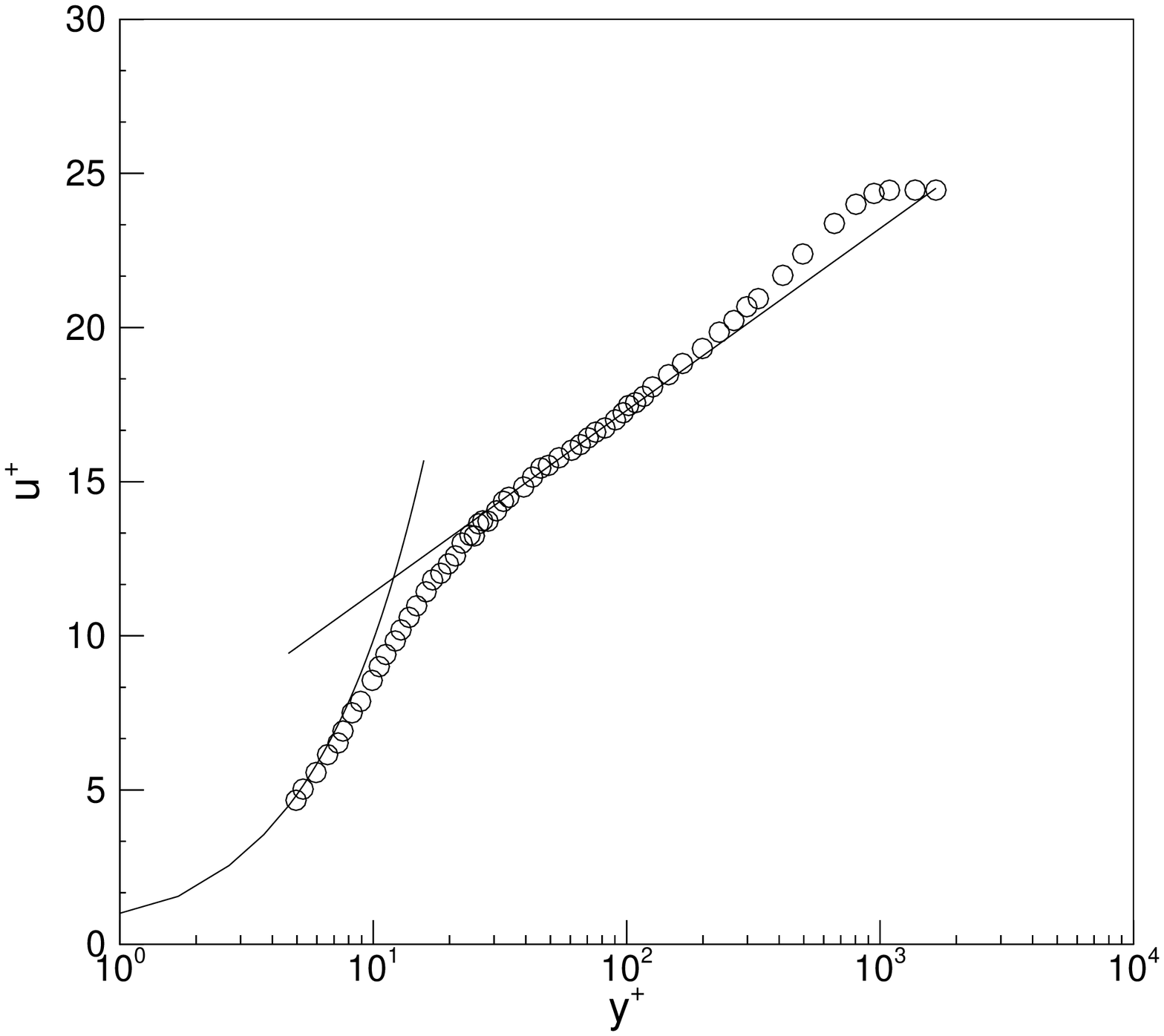,width=7.5cm} }
   \centerline{ \epsfig{figure=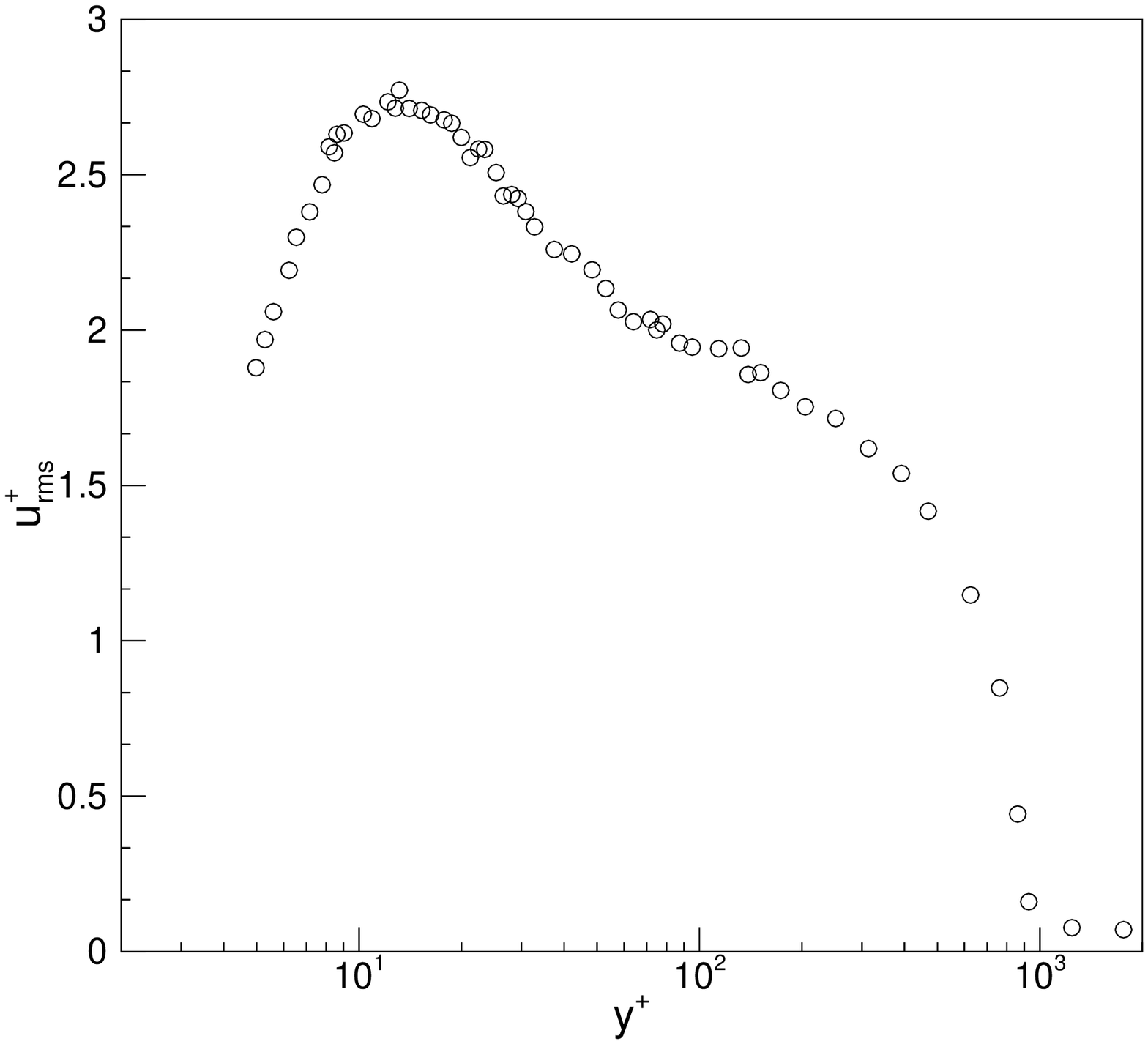,width=7.5cm} }
   \vspace*{2.cm}
   \caption{}
\end{figure}
\newpage
\begin{figure}[b!]
   \vspace*{2.cm}
   \centerline{ \epsfig{figure=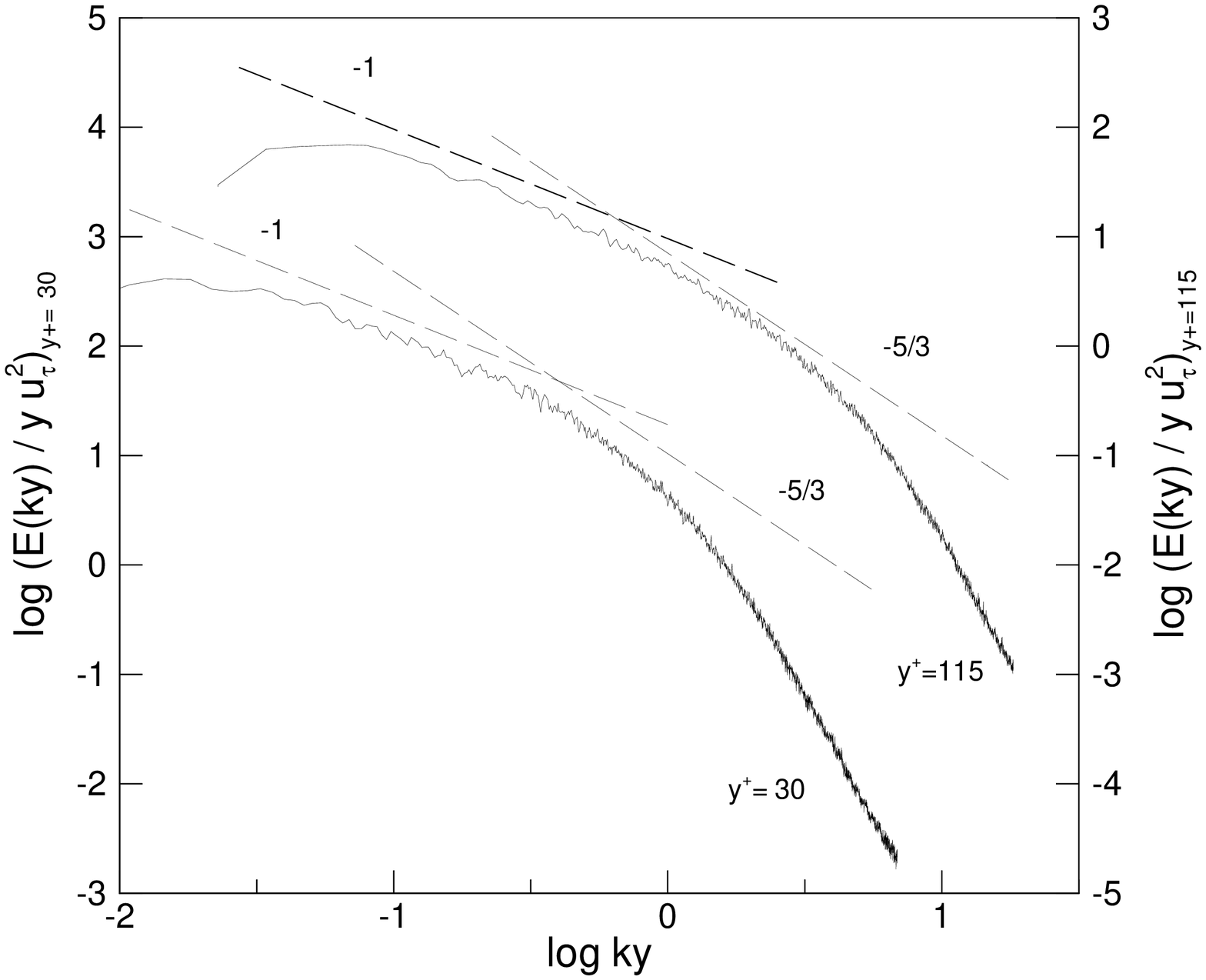,width=7.5cm} }
   \centerline{ \epsfig{figure=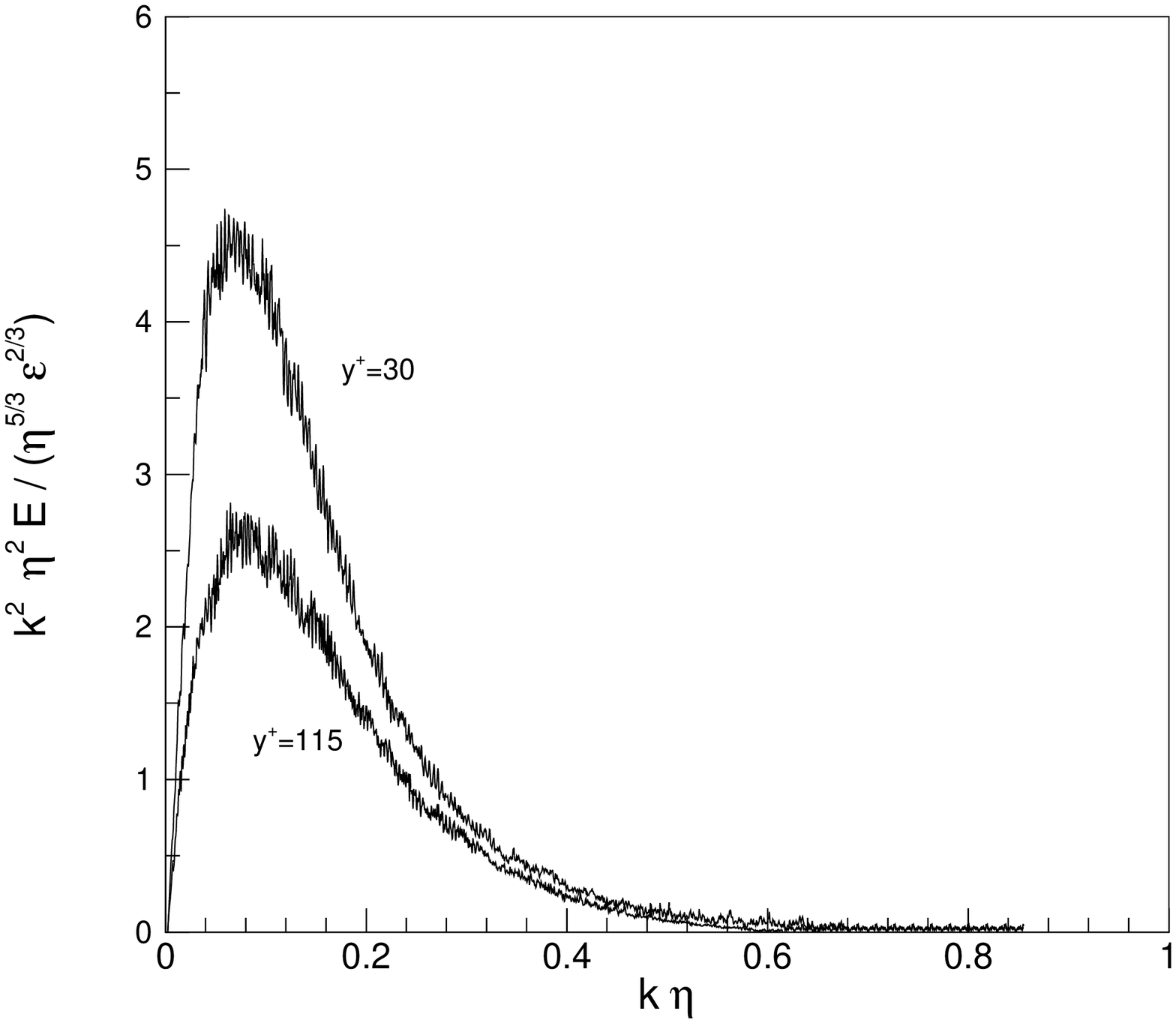,width=7.5cm} }
   \vspace*{2.cm}
   \caption{}
\end{figure}
\newpage
\begin{figure}[t]
   \vspace*{2.cm}
   \centerline{ \epsfig{figure=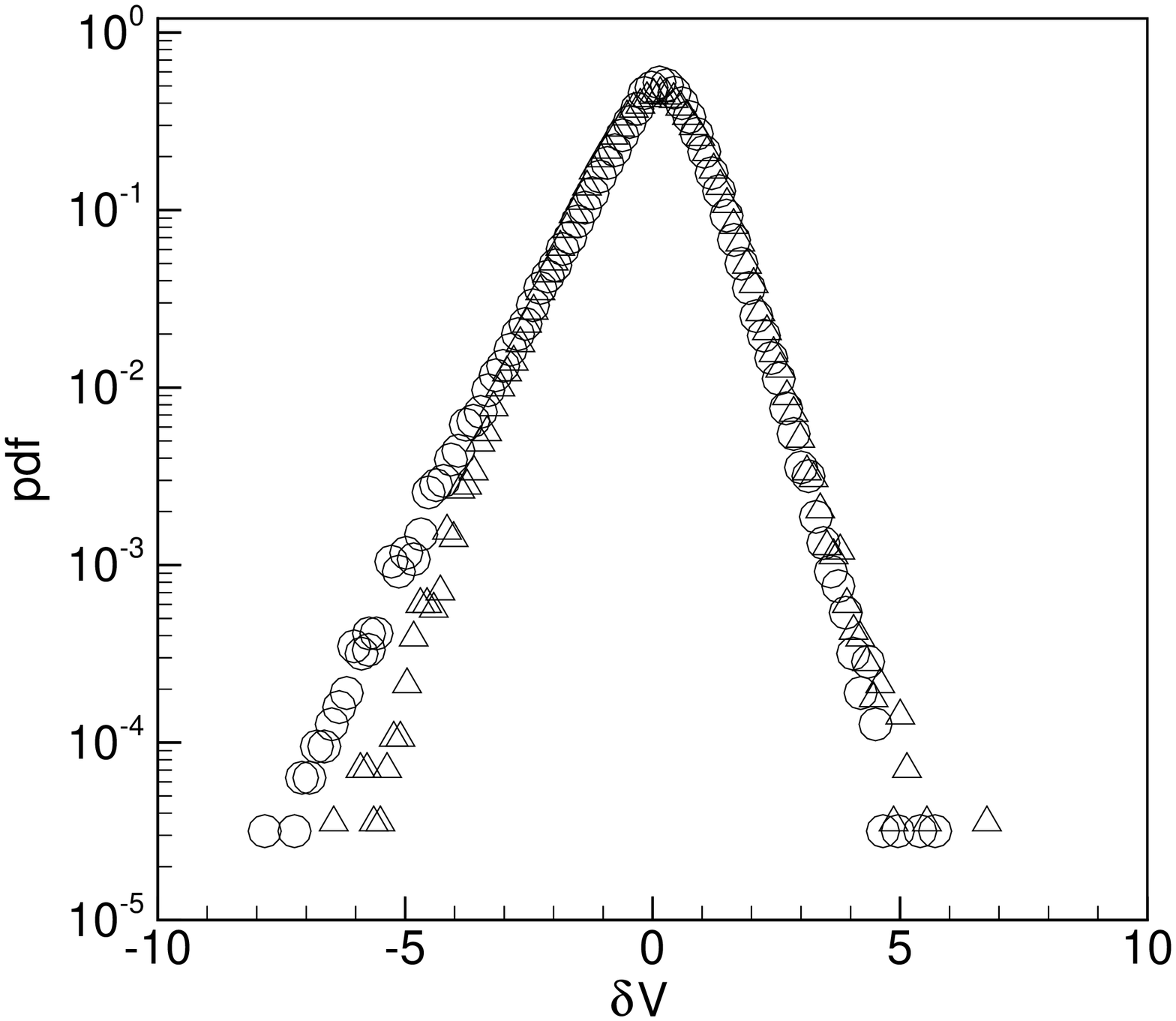,width=6.cm} }
   \centerline{ \epsfig{figure=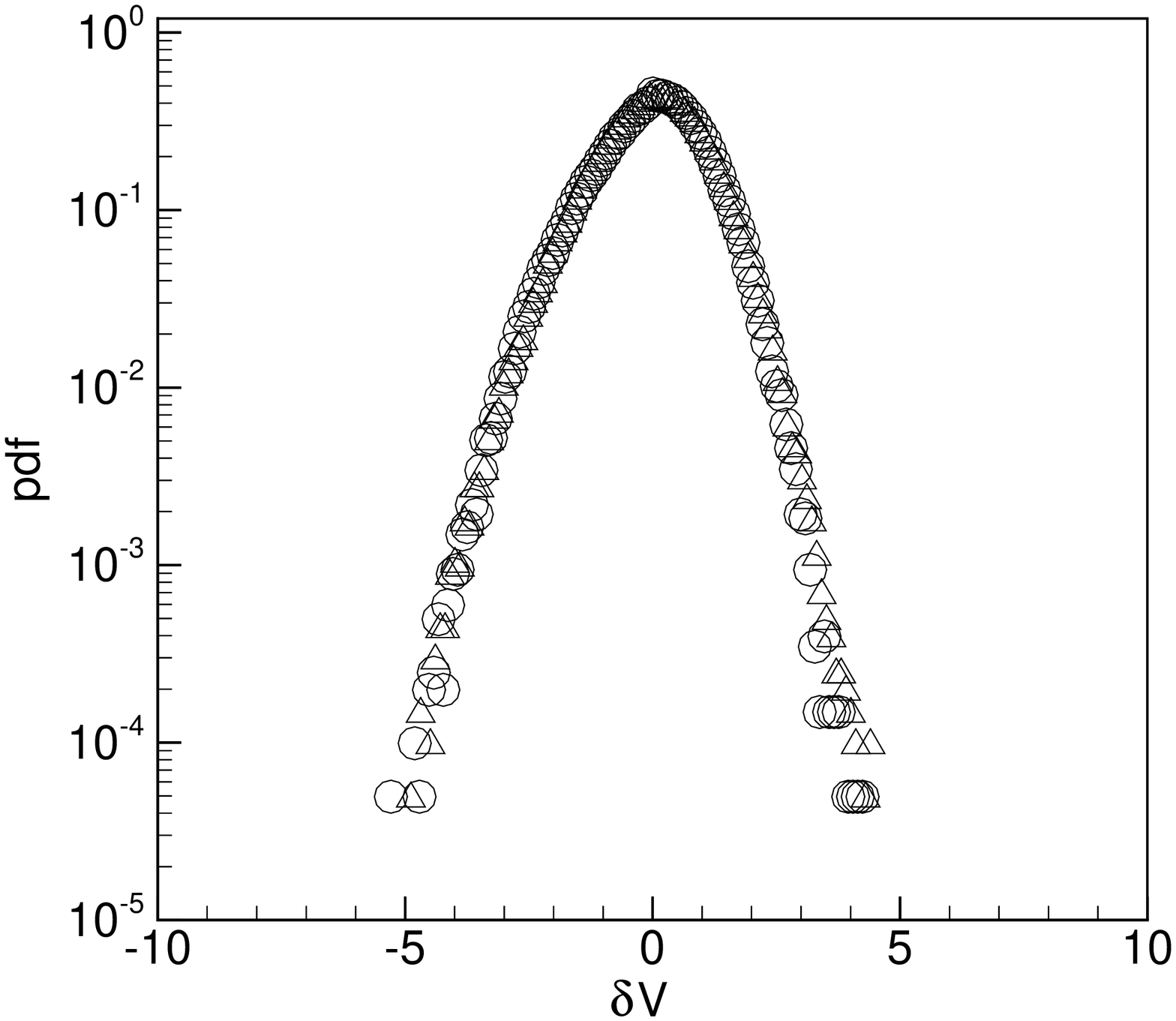,width=6.cm} }
   \vspace*{2.cm}
   \caption{}
\end{figure}
\newpage
\begin{figure}[t]
   \vspace*{2.cm}
   \centerline{ \epsfig{figure=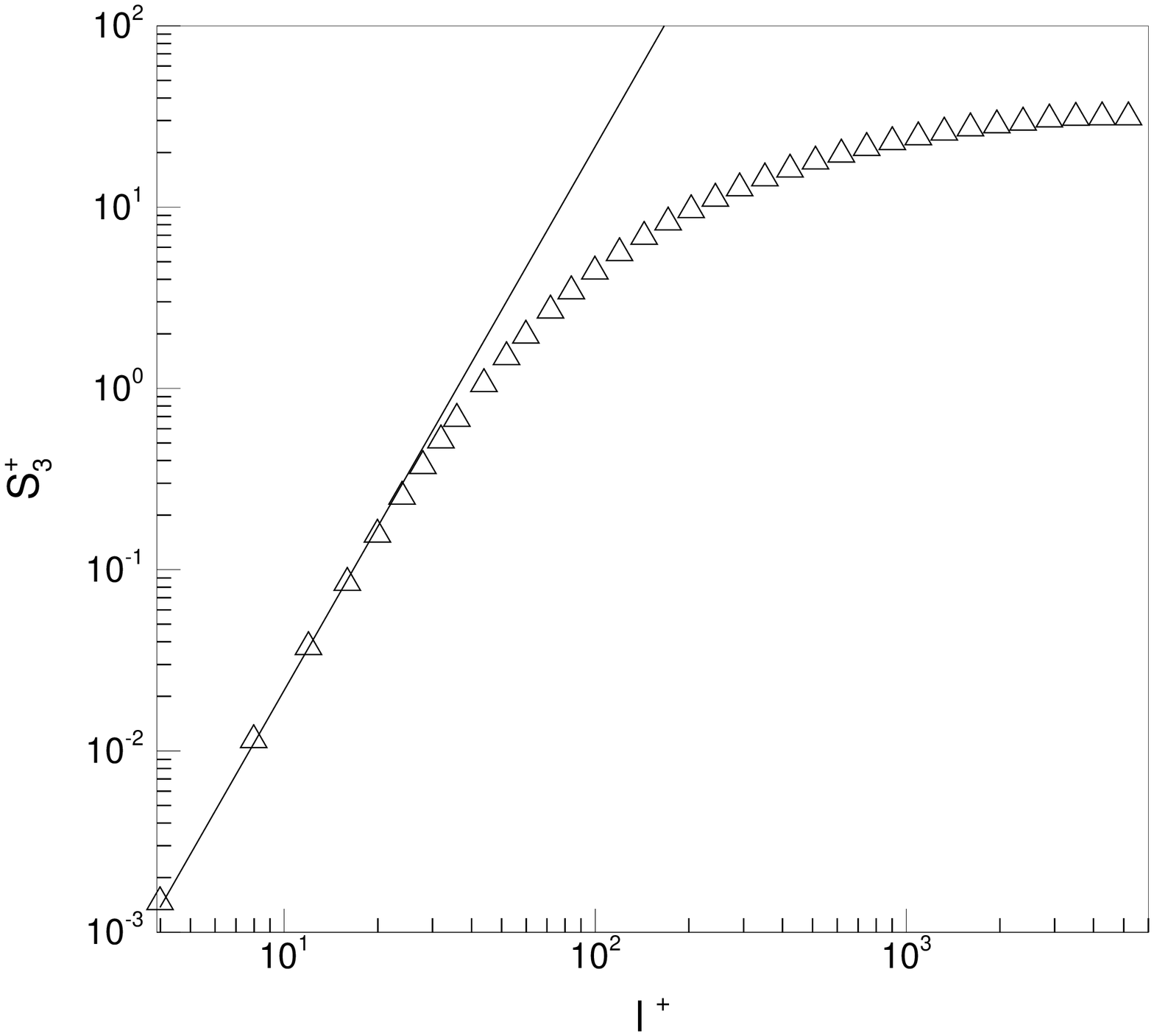,width=5.cm} }
   \centerline{ \epsfig{figure=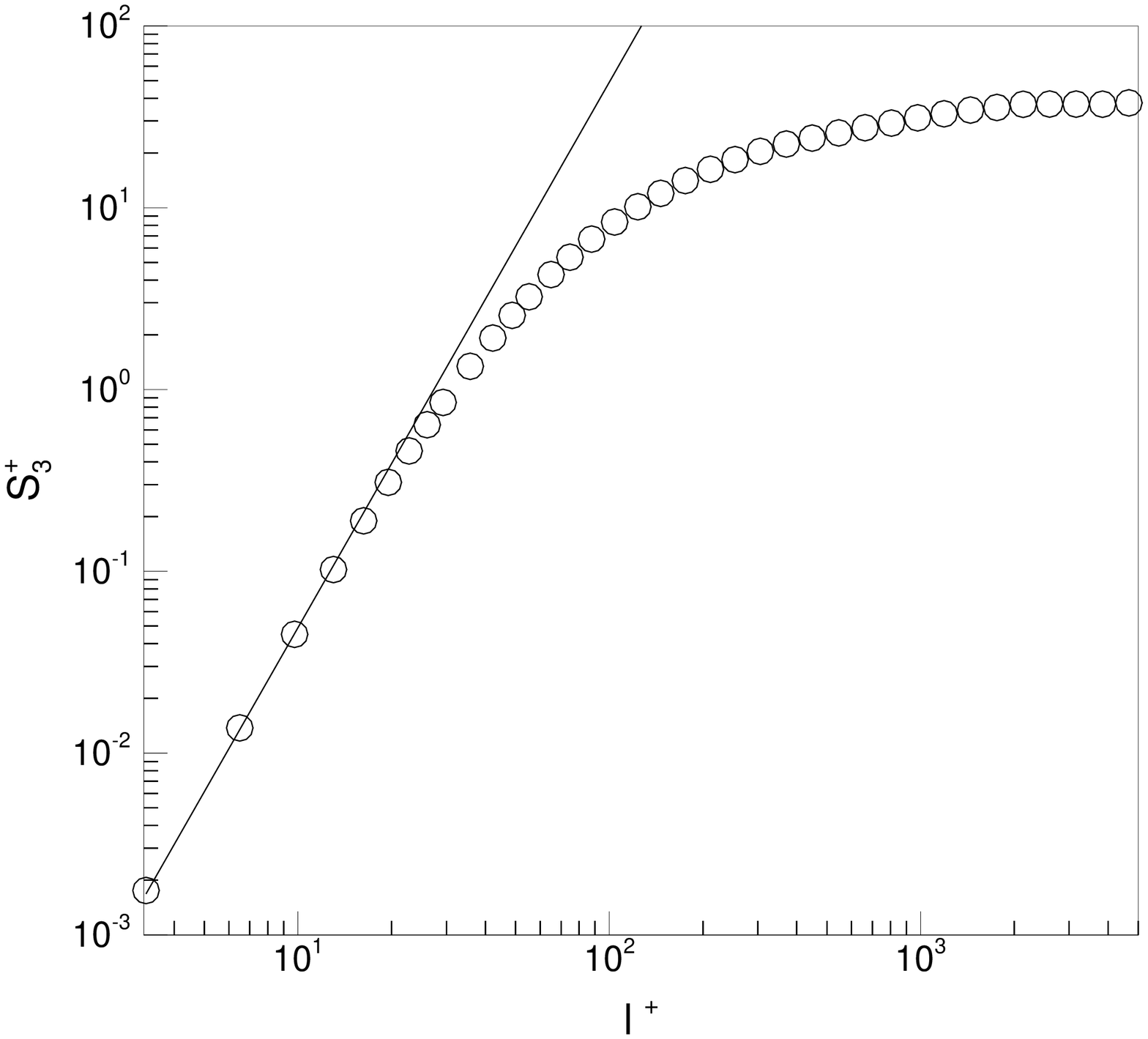,width=5.cm} }
   \vspace*{2.cm}
   \caption{}
\end{figure}
\newpage
\begin{figure}[t]
   \vspace*{2.cm}
   \centerline{ \epsfig{figure=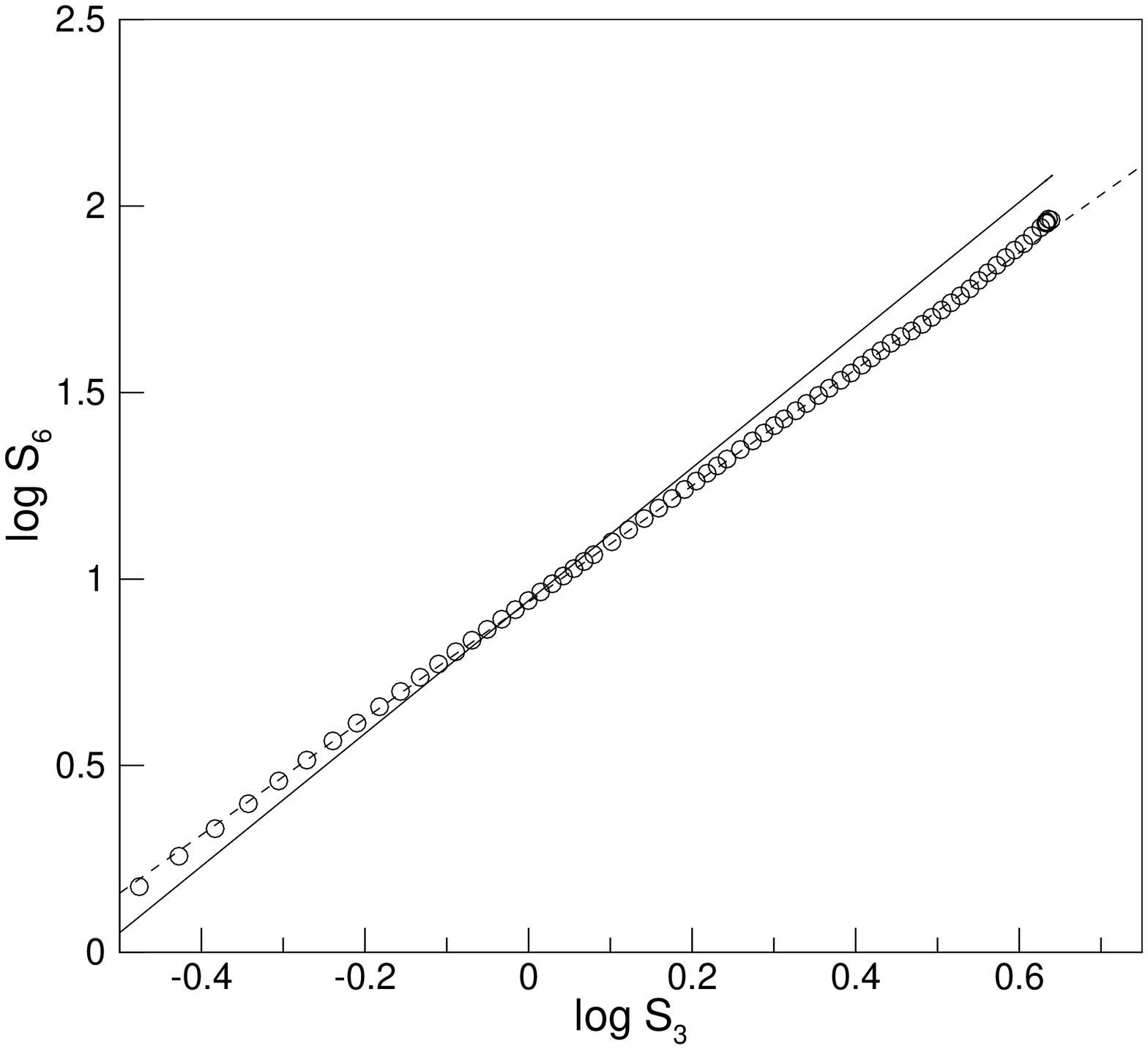,width=6.5cm} }
   \centerline{ \epsfig{figure=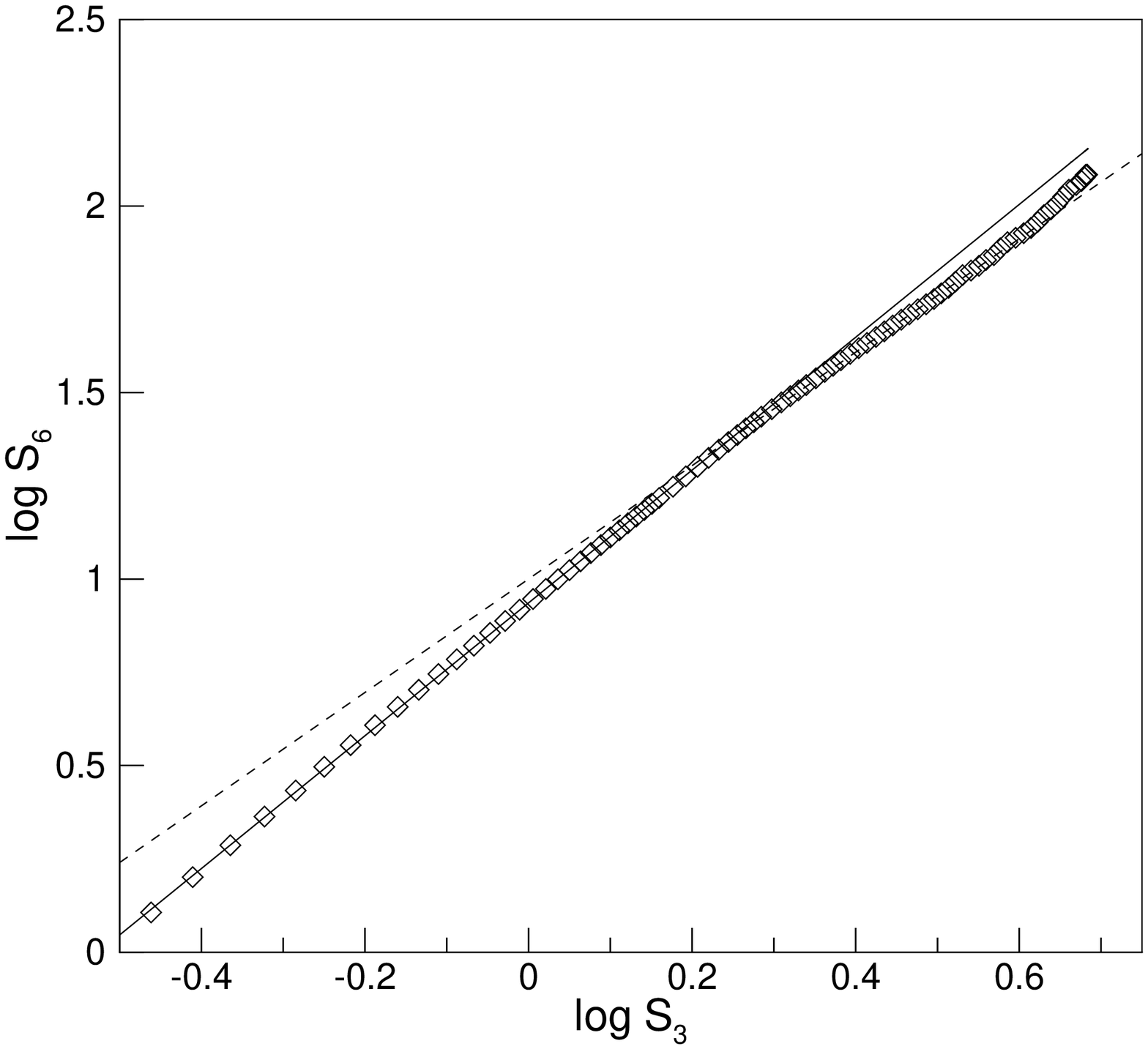,width=6.5cm} }
   \centerline{ \epsfig{figure=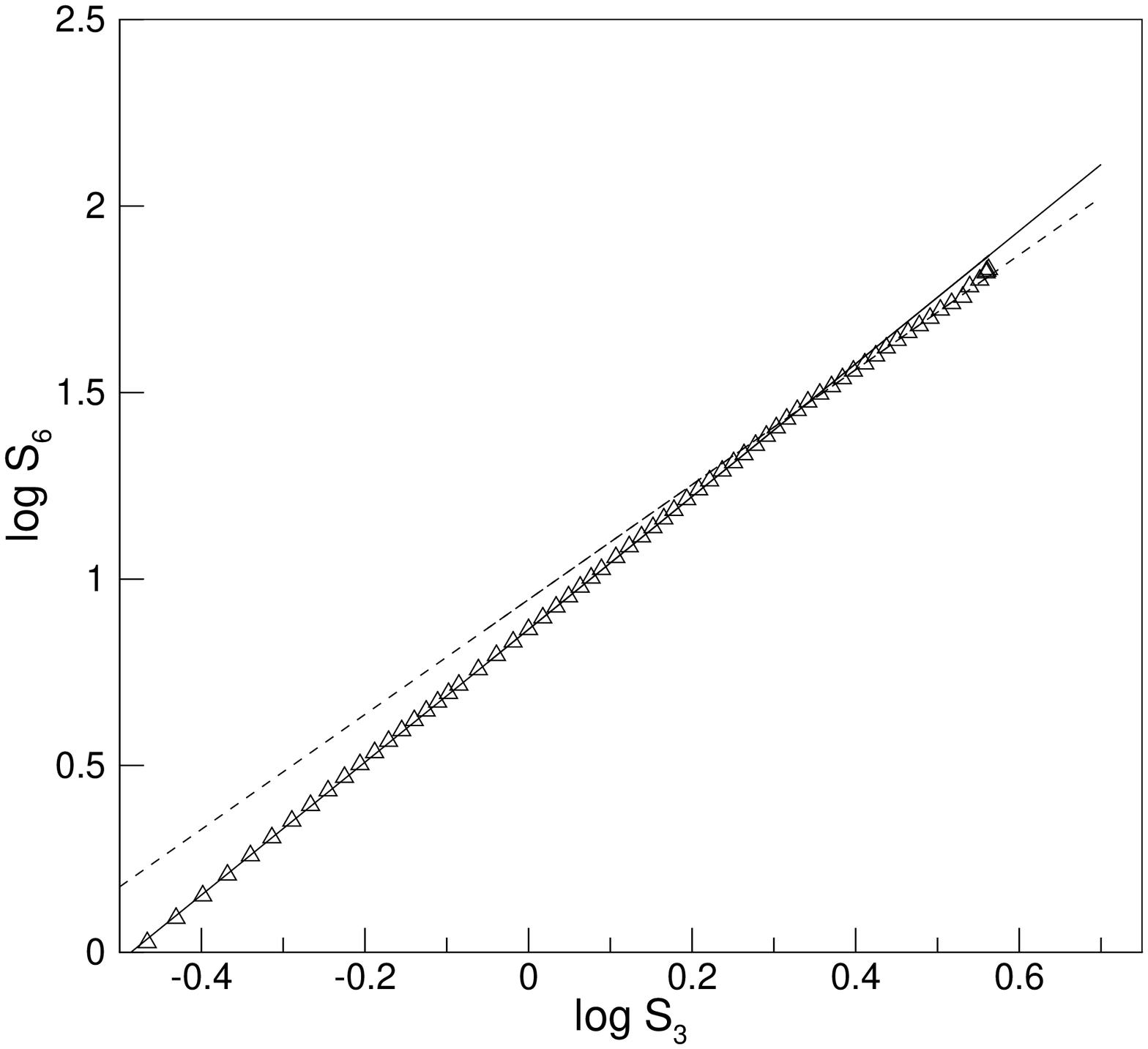, width=6.5cm} }
   \vspace*{2.cm}
   \caption{}
\end{figure}
\newpage
\begin{figure}[t!]
   \vspace*{2.cm}
   \centerline{ \epsfig{figure=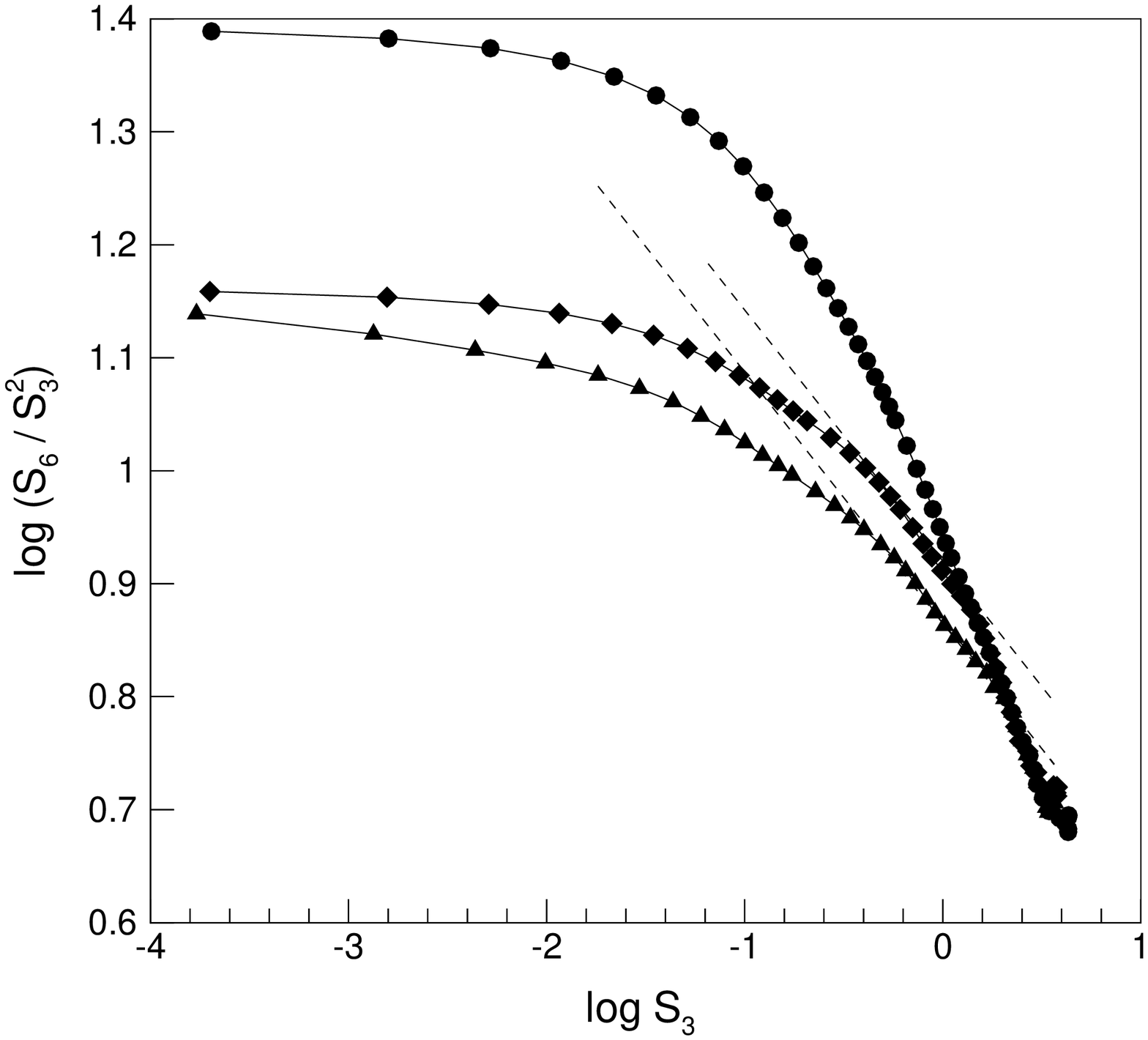,width=7.5cm} }
   \centerline{ \epsfig{figure=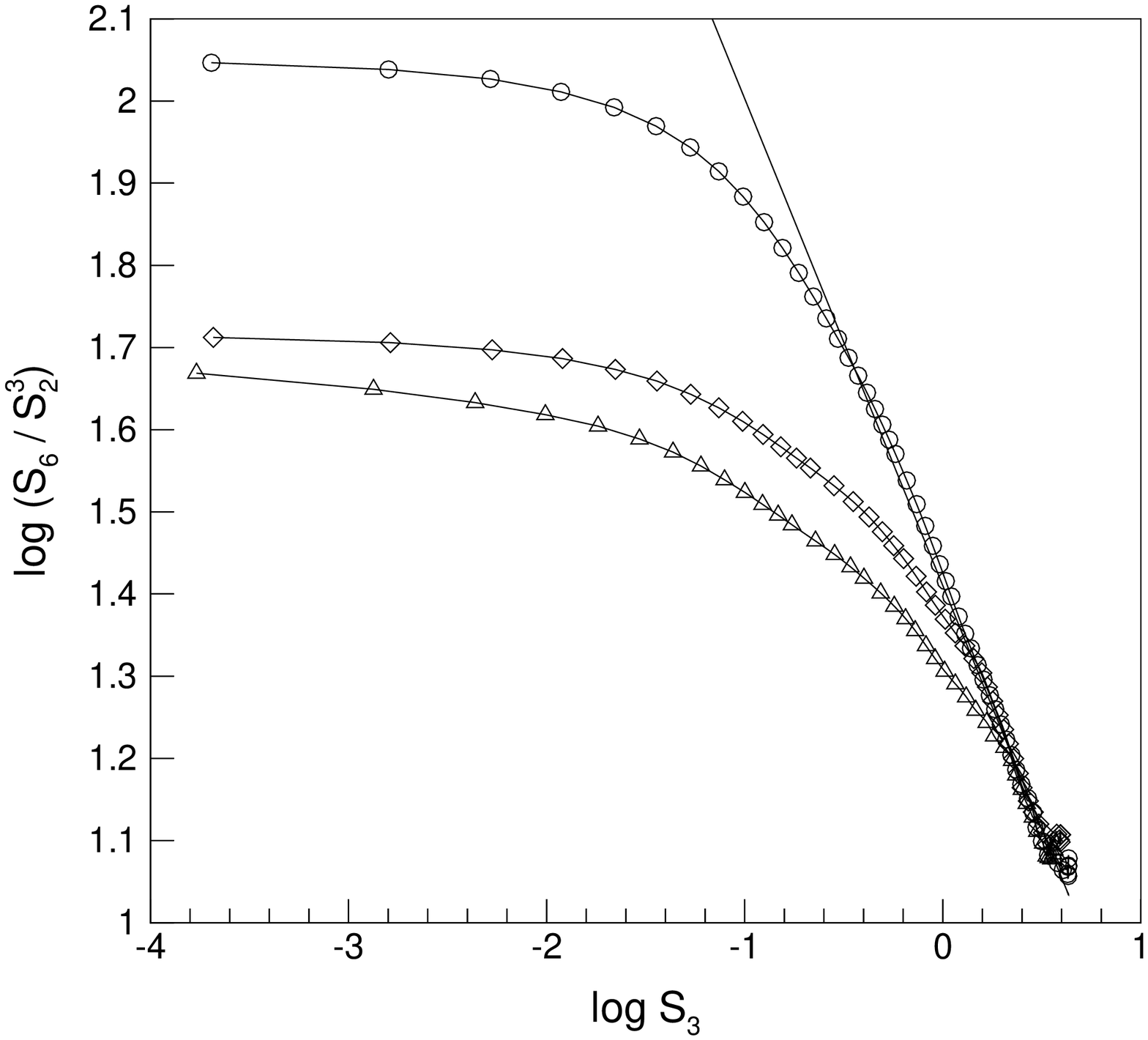,width=7.5cm} }
   \vspace*{2.cm}
   \caption{}
\end{figure}
\newpage
\begin{figure}[b!]
   \vspace*{1.cm}
   \centerline{ \epsfig{figure=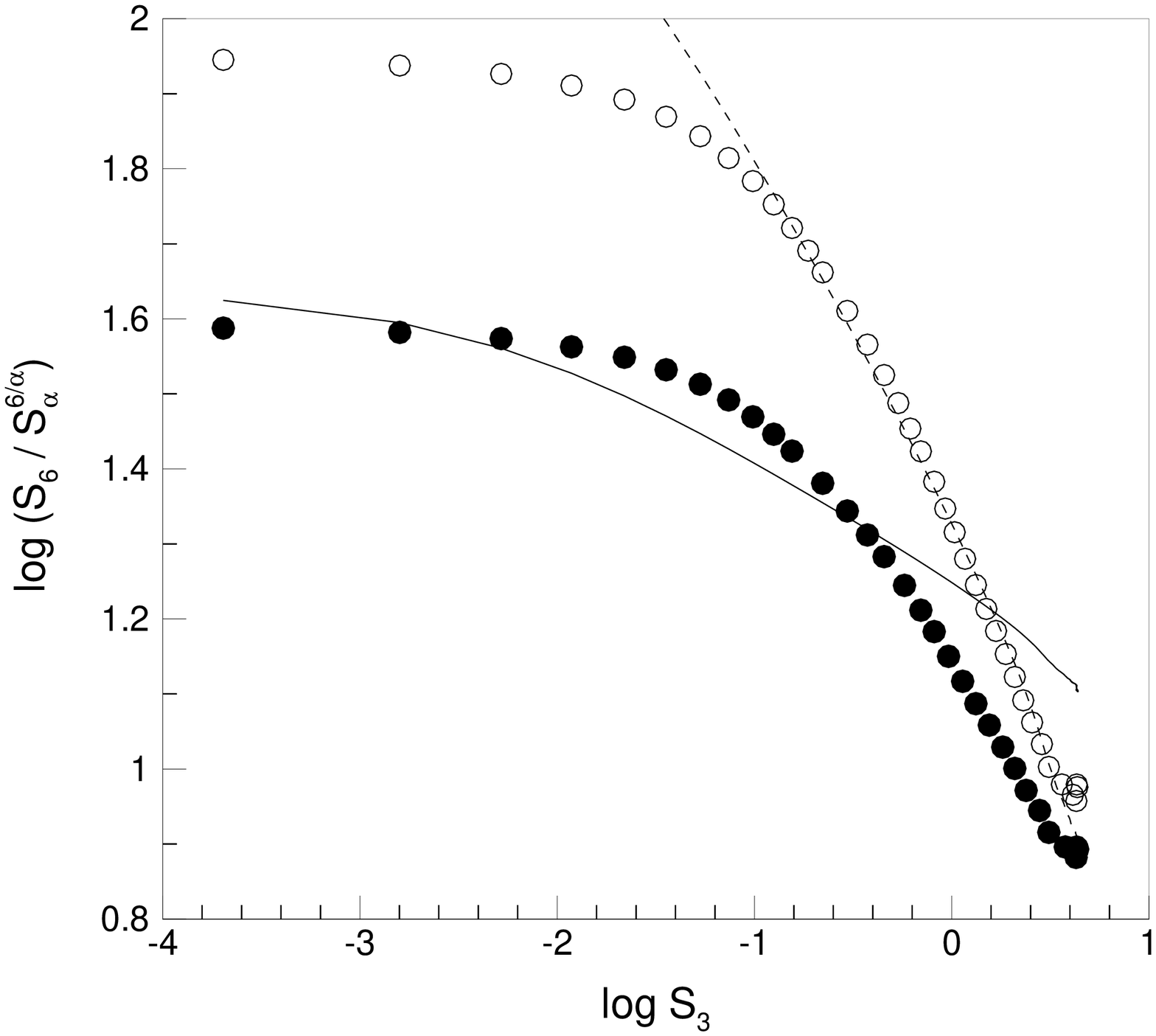, width=7.cm} }
   \centerline{ \epsfig{figure=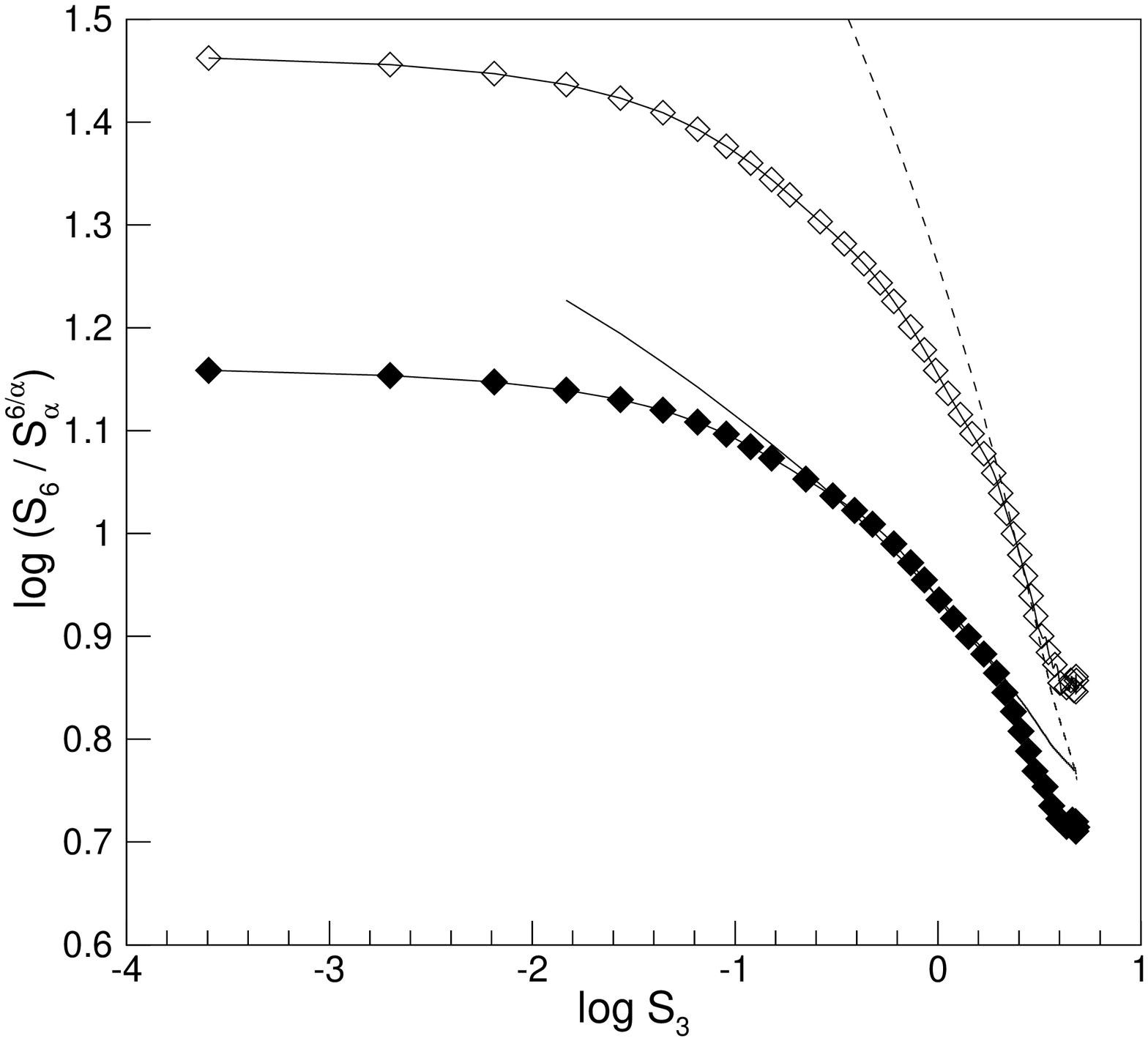,width=7.cm} }
   \centerline{ \epsfig{figure=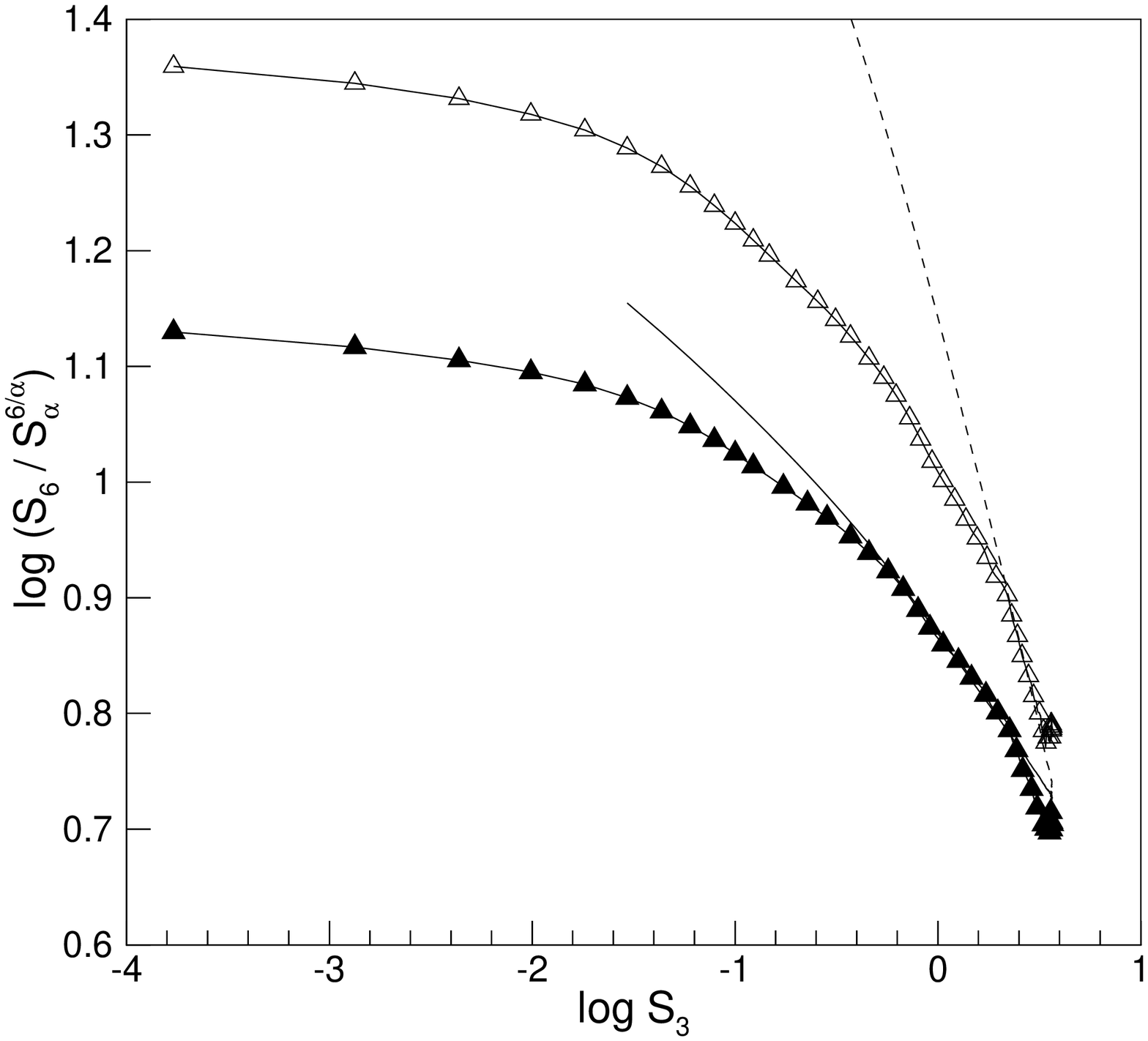, width=7.cm} }
   \vspace*{1.cm}
   \caption{}
\end{figure}
\newpage
\begin{figure}[b!]
   \vspace*{2.cm}
   \centerline{ \epsfig{figure=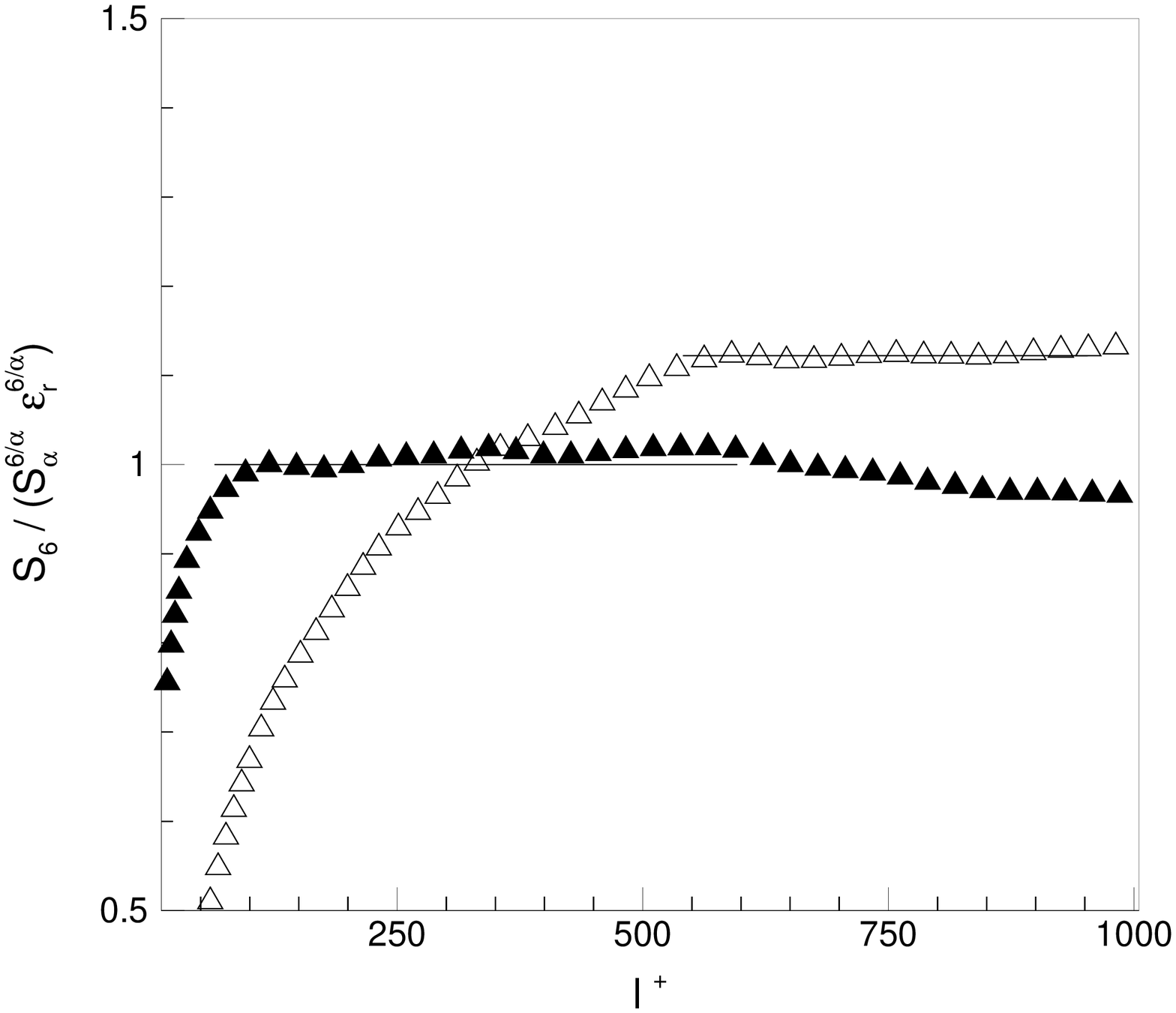,width=6.cm} }
   \centerline{ \epsfig{figure=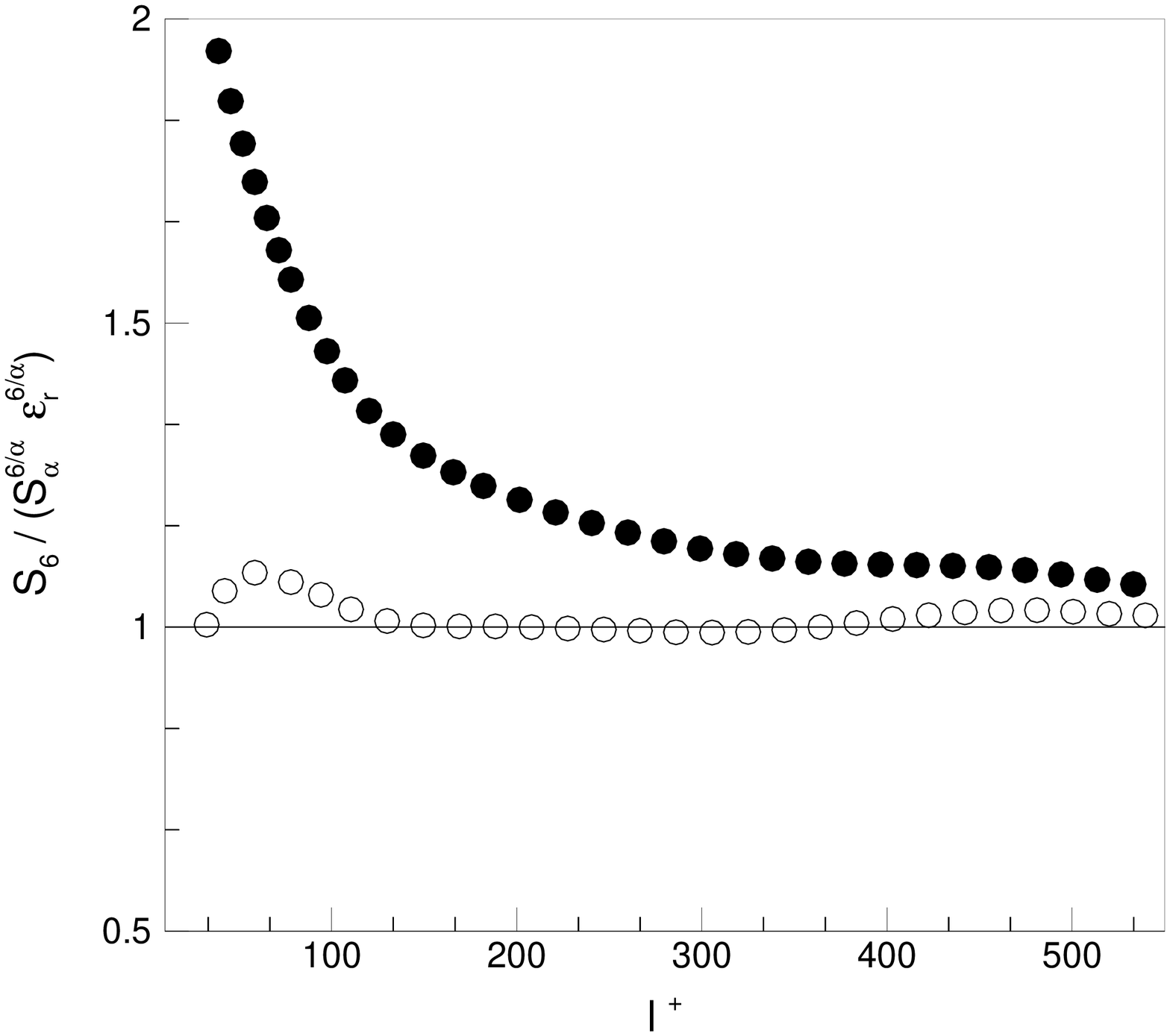,width=6.cm} }
   \vspace*{2.cm}
   \caption{}
\end{figure}
\newpage
\begin{figure}[t!]
   \vspace*{2.cm}
   \centerline{ \epsfig{figure=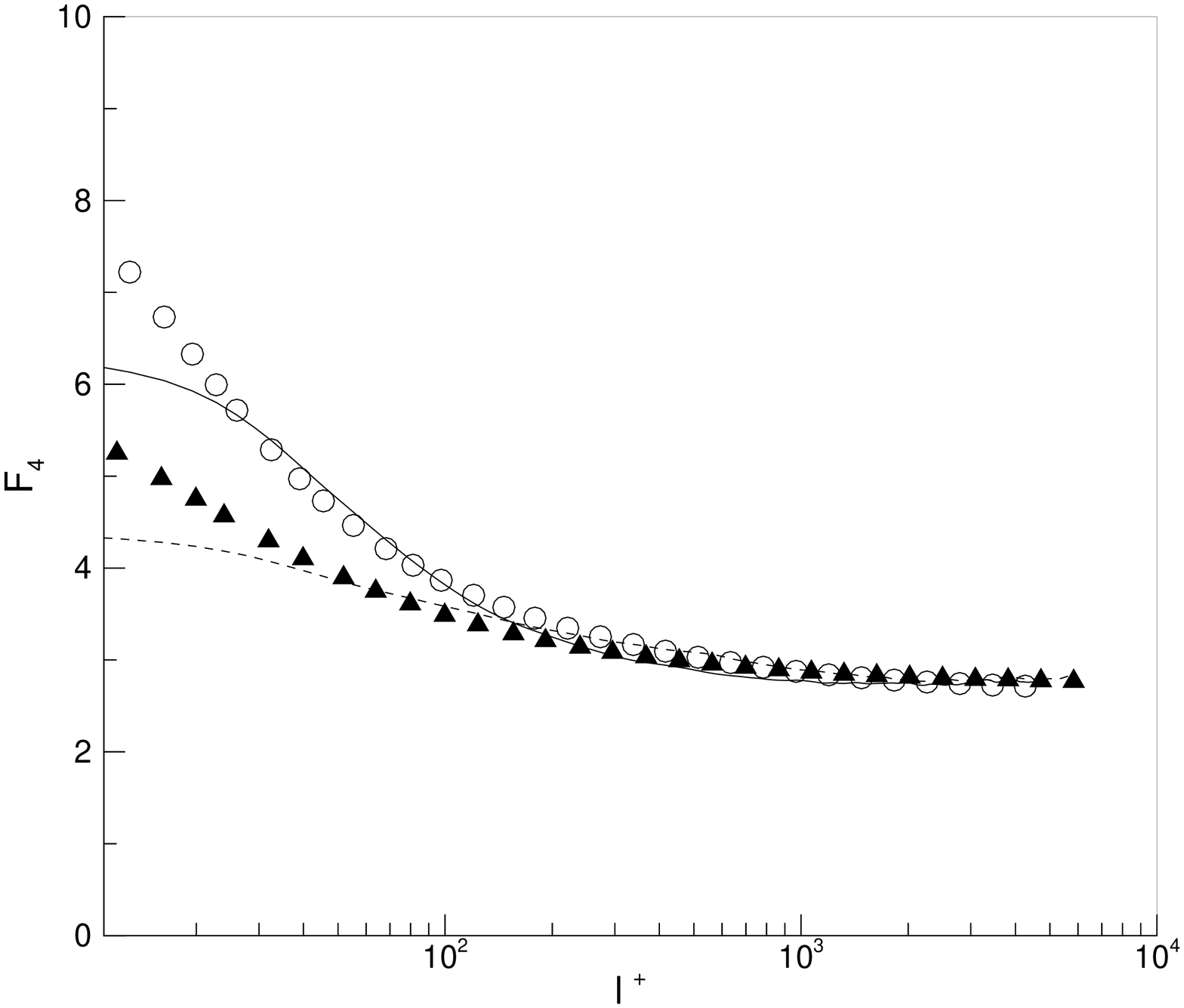,width=9.cm} }
   \vspace*{2.cm}
   \caption{}
\end{figure}
\newpage
\begin{table}[b]
\vspace*{2.cm}
\begin{center}
\begin{tabular}{r|ccccccccc}
$y^+$&$u_{rms}$&$\bar \epsilon$&$\lambda$&$\eta$&$L_s$&$l_d$&$Re_{\lambda}$&$S^*$ &$S_c^*$\\
\hline
$115$ & 0.96 & 31.7  &  2.8   &0.102 & 1.24  & 27.9 & 179 &  8.  &0.19 \\
$ 70$ & 1.02 & 44.2  &  2.5   &0.093 & 0.67  & 22.2 & 170 & 10.3 &0.27 \\
$ 30$ & 1.21 & 48.7  &  2.5   &0.091 & 0.17  & 36.4 & 202 & 36.1 &0.67 \\ 
\hline
units&${\rm m/s}$&${\rm m^2/s^3}$&${\rm mm}$&${\rm mm}$&${\rm mm}$&${\rm mm}$&/&/&/ 
\end{tabular}
\vspace*{2.cm}
\caption{Main turbulent parameters at the three measurement points. 
\label{table}}
\end{center}
\end{table}
\newpage

\end{document}